\documentclass[
 amsmath,amssymb,
 aps,
]{revtex4-2}

\usepackage{graphicx}% Include figure files
\usepackage{dcolumn}% Align table columns on decimal point
\usepackage{bm}% bold math
\usepackage[]{graphicx,color}
\usepackage[makeroom]{cancel}
\usepackage[position=t,caption=false,justification=centering]{subfig}

\usepackage[position=t,caption=false,justification=centering]{subfig}
\usepackage{mathtools}
\renewcommand{\vec}[1]{{\mbox{\boldmath $ #1 $}}}

\newcommand{\Cu}{\mbox{Cu}}  % Carreau number
\newcommand{\Rey}{\mbox{Re}} % Reynolds number

\graphicspath{{./figures/}}

\begin{document}

\preprint{APS/123-QED}

\title{Flow of a shear-thinning fluid in a rectangular duct}

\author{Ilya Barmak}
\email{ilyab@tauex.tau.ac.il}
\affiliation{School of Mechanical Engineering, Tel Aviv University, Tel Aviv 6997801, Israel}
\affiliation{Soreq NRC, Yavne 8180000, Israel 
}%

\author{Davide Picchi}%
\email{davide.picchi@unibs.it}
\affiliation{Department of Mechanical and Industrial Engineering, Università degli Studi di Brescia, 
via Branze 38, Brescia 25123, Italy 
}%

\author{Alexander Gelfgat}
\email{gelfgat@tauex.tau.ac.il}
\affiliation{School of Mechanical Engineering, Tel Aviv University, Tel Aviv 6997801, Israel 
}%

\author{Neima Brauner}
\email{brauner@tauex.tau.ac.il}
\affiliation{School of Mechanical Engineering, Tel Aviv University, Tel Aviv 6997801, Israel 
}%

%\date{\today}

\begin{abstract}
We address the problem of steady laminar flow of a shear-thinning fluid in a rectangular duct, which is encountered in many systems, in particular, in microfluidic and biomedical devices. However, an exact solution for the flow of non-Newtonian fluids that considers a realistic shear-thinning rheological behavior is not available in the literature. In this study, an accurate solution for the case of Carreau fluid is obtained and investigated numerically. The numerical solution allows us to analyze the effects of the fluid rheology and the aspect ratio of the rectangular duct on the velocity field and pressure gradient that drives the flow. The relationship between the pressure gradient and the Carreau number is found to follow the rheological curve of the shear-thinning fluid. The analysis shows that the fluid rheology and the aspect ratio have independent contributions to the integral flow characteristics. Moreover, separate consideration of these contributions allows us to arrive at universal scaling and general formulae for the pressure gradient and friction factor for various rheological parameters of the fluid and aspect ratios of rectangular ducts.
\end{abstract}

\maketitle

%%%%%%%%%%%%%%%%%%%%%%%%%%%%%%%%%%%%%%%%%%%%%%%%%%%%%%%%%%%
\section{Introduction} \label{Sec: Introduction}

The flow of a shear-thinning fluid in confined environments is typical for many biological and engineering systems. Examples include polymer extrusions \cite{Bird87}, microfluidic and biomedical devices \cite{Chhabra07,Secomb17}, or flows in fractures and porous material \cite{Bear88}.  
In these contexts, the fluid moves through arbitrary channels whose cross-section rarely can be idealized with pipe or planar geometries.

A typical set-up is a rectangular duct with a finite height-to-width aspect ratio. In this setting, the flow is driven by a pressure gradient and the cross-sectional geometry determines both the resistance to flow and the shape of the velocity profile. A number of studies in the literature have focused on Newtonian fluids (e.g., \cite{Tachibana81,Spiga94,Xing13}), starting from the seminal work of \citet{Cornish28} that provided the exact solution for pressure gradient dependence on the flow rate in the case of laminar fully developed flow in a straight rectangular duct. 

In many practical scenarios, the working fluids (e.g., polymers solutions, suspensions, emulsions and biological fluids) exhibit a shear-thinning behavior and their viscosity is a function of the imposed shear rate. At low and high shear rates, the viscosity attains constant values, commonly known as the zero-shear-rate and infinite-shear-rate viscosity,
respectively, and  the fluid behavior is Newtonian. On the other hand, at intermediate shear rates, the viscosity decreases with increasing the shear rate following a power-law behavior \cite{Bird87}. 

Among the rheological models available in the literature, only a few can properly describe the fluid rheology in the entire range of shear-rates, in particular, the Carreau–Yasuda \cite{Carreau72,Yasuda81} and the Ellis models \cite{Reiner65}. However, due to the fact that the effective viscosity is a nonlinear function of the shear-rate, an analytical solution for the velocity profile of a shear-thinning fluid flowing in a rectangular channel with arbitrary aspect ratio is still missing in the literature. 

Analytical solutions available in the literature for Ellis and power-law fluids \cite{Bird87} are only for simple geometries, e.g. planar and circular channels, where the flow is assumed to be one dimensional. More recently, \citet{Sochi15} proposed a semi-analytical methods to compute the flow rate curves for Carreau fluids in a one-dimensional geometry, following the  
Weissenberg–Rabinowitsch–Mooney methodology, that requires a numerical computation of integrals expressed in terms of hyper-geometric functions. \citet{Boyko21} proposed a theoretical approach to calculate the flow rate-pressure gradient relation for Carreau fluids in a narrow channel (i.e., where the height of the channel is much smaller compared to its width) by deriving the asymptotic solutions for the small, intermediate, and high shear rate regions. Their theoretical predictions have been compared with the experimental data of \citet{Pipe08}, where the test fluid is a xanthan gum solution flowing in a slit microchannel. \citet{Sun21} investigated the flow of a Carreau fluid showing a yield-stress in a circular pipe geometry exploring the effect of the fluid rheology on the pressure gradient. An analytical solution for a Carreau fluid flowing between two parallel plate was obtained by \citet{Griffiths20} only for fluids whose shear-thinning index is restricted to a discrete set of values. Otherwise, no exact solution can be derived.

When the channel shape is more complex, as in case of a rectangular duct, the flow field should be obtained numerically. Researchers have focused primarily on the steady flow of power-law fluids using different numerical methods, such as Galerkin \cite{Schechter61} or finite element methods \cite{Syrjala95,Palit72}. Unfortunately, the power-law model is not capable of capturing the plateau in viscosity at small and very high shear rates. According to this model, the viscosity attains an unrealistic unbounded growth at low shear rates and a contentious decrease at high shear rates. This leads to significant errors in the prediction of the local velocity field and integral flow characteristics \citep[e.g., ][]{Picchi17,Picchi21,Shahsavari15,Boyko21}. Since the flow field in channels with a finite aspect ratio is rather complex (e.g., flow in the corners), a proper description of flow of realistic shear-thinning fluids must not ignore those limiting behaviors. 

To fill those gaps, the goal of this paper is to study the flow characteristics of a shear-thinning (Carreau) fluid though a rectangular channel with an arbitrary aspect ratio. We aim at (i) clarifying the competition between the zero-shear-rate, infinite-shear-rate,  and the shear-thinning effects on the velocity field and the flow resistance due to the shear-stress acting on the solid boundaries; and (ii) understanding the impact of the channel aspect ratio on flow characteristics in the context of shear-thinning fluids. We aspire at assessing the contribution of both the fluid rheology and the duct aspect ratio to generalize the classical friction factor formula for shear-thinning fluids. This becomes possible after identification of a universal scaling law for the effective viscosity, which is a function of the problem dimensionless parameters. The generalization of the friction factor applies to both Newtonian and inelastic shear-thinning fluids flowing in a rectangular duct of an arbitrary aspect ratio.  

To this aim, the problem is formulated in dimensionless coordinates (Section \ref{Sec: Formulation}). The steady laminar flow of a shear-thinning fluid, whose viscosity is described by the Carreau model, is solved numerically (Section \ref{Sec: Numerics}). The results are validated by comparison with the available literature data in Section \ref{Sec: Results_a}, where it is shown that the results converge to the classical results of \citet{Cornish28} in the Newtonian limit. The effect of the fluid rheology and the aspect ratio on the velocity field and pressure gradient is discussed in detail in Sections \ref{Sec: Results_a} and  \ref{Sec: Results_b}, respectively. Finally, we show how the universal scaling law and generalized formulae can be derived for the pressure gradient and friction factor of a realistic shear-thinning fluid flowing in rectangular ducts (Section \ref{Sec: Results_c}).

%%%%%%%%%%%%%%%%%%%%%%%%%%%%%%%%%%%%%%%%%%%%%%%%%%%%%%%%%%%
\section{Problem formulation} \label{Sec: Formulation}

We consider a steady and fully developed laminar flow of a non-Newtonian shear-thinning liquid in a rectangular duct of height $H$, width $W$, and cross-sectional area as $A = H W$. The height-to-width aspect ratio of the duct is defined as $\displaystyle\varepsilon=H/W$. Under these conditions, the flow is unidirectional. The only non-zero velocity component is in the axial direction, which varies only in the flow cross section $\vec{u}=(0,0,u(x,y))$. The dimensionless momentum equation reads:
\begin{equation} \label{Eq: Momentum}
	\frac{\partial \tau_{x z}}{\partial x}
	+ \frac{\partial \tau_{y z}}{\partial y}
	= \frac{d p}{d z},
\end{equation}
where $\displaystyle\tau_{x z}$ and $\displaystyle\tau_{y z}$ are shear stresses in the duct cross-section and $\displaystyle dp/dz$ is the pressure gradient, which is constant across the duct cross-section. 
The momentum equation is normalized using the fluid average velocity, i.e., the flow rate divided by the cross-sectional area ($\displaystyle U = Q/A$), as a 
scale for velocity. Considering the duct height $H$ as the relevant length scale and the viscous force based on the zero-shear-rate viscosity, $\displaystyle\mu_0 U/H^2$ is the scale for the pressure gradient. 

In this work, the fluid viscosity is described by the Carreau \cite{Carreau72} rheological model due to its capability of capturing both the low- and high-shear rates constant viscosity behaviors. The dimensionless constitutive relations of a Carreau fluid for a unidirectional flow are:
\begin{equation}
	\tau_{x z} = \mu \big(\dot{\gamma}\big) 
                    \dot{\gamma}_{x z},
	\quad
	\tau_{y z} = \mu \big(\dot{\gamma}\big) 
                     \dot{\gamma}_{y z},
	\quad
	\dot{\gamma}_{x z} = \frac{\partial u}{\partial x},
	\quad
	\dot{\gamma}_{y z} = \frac{\partial u}{\partial y},
        \quad
        \dot{\gamma} = \sqrt{\dot{\gamma}_{x z}^2 + \dot{\gamma}_{y z}^2},
\end{equation}
where $\displaystyle\dot{\gamma}_{x z}$, $\displaystyle\dot{\gamma}_{y z}$, and $\displaystyle\dot{\gamma}$ are the non-zero components of strain tensor and its norm.
The  dimensionless viscosity (scaled by $\mu_0$) is given by
\begin{equation} \label{Eq: Mu}
	\mu 
	= m 
	+ \bigg(1 - m\bigg)
	\Biggl(1
	+ \Cu^2
	\bigg[
            \bigg(\frac{\partial u}{\partial x}\bigg)^2
		+ \bigg(\frac{\partial u}{\partial y}\bigg)^2
       \bigg]
	 \Biggr)^{(n-1)/2},
\end{equation}
where $\displaystyle m = \mu_\infty/\mu_0$ is the ratio of infinity- to zero-shear-rate dynamic viscosities and $n$ the shear-thinning index. The additional dimensionless parameter in Eq. (\ref{Eq: Mu}) is the Carreau number defined as 
\begin{equation}\label{Eq: Cu}
    \Cu = \frac{\lambda U}{ H},
\end{equation}
where $\lambda$ is the time constant of the non-Newtonian liquid.  Specifically, the Carreau number controls the onset of the shear-thinning effect:  when $\Cu \to 0$ the effective viscosity reduces to the zero-shear-rate Newtonian limit ($\mu=1$), while for $\Cu \to \infty$, the infinity-shear-rate effect dominates and the dimensionless viscosity is equal to $m$.  At intermediate shear rates the rheology is dominated by the shear-thinning effect and the viscosity decreases following the typical power-law behavior.  

Substituting the expression for the effective viscosity, Eq.\ (\ref{Eq: Mu}), into the momentum equation, Eq.\ (\ref{Eq: Momentum}), the governing equation for the velocity reads
\begin{equation}  \label{Eq: Momentum_Carreau}
    \frac{\partial}{\partial x} \bigg(\mu \frac{\partial u}{\partial x}\bigg)
	+ \frac{\partial}{\partial y} \bigg(\mu \frac{\partial u}{\partial y}
	\bigg)
	= 
	\frac{d p}{d z}.
\end{equation}
The velocity field and the unknown pressure gradient can be found by combining Eq.\ (\ref{Eq: Momentum_Carreau}) with the flow rate constraint, i.e., the velocity in the flow cross section provides the (known) fluid flow rate, $Q$. This results in the following dimensionless relation:
\begin{equation} \label{Eq: Flow_rate_constraint}
    \varepsilon \int_{0}^1 \int_{0}^{1/\varepsilon} u\, dx\, dy = 1.
\end{equation}

Differently from the case of Newtonian fluid, where the solution of the linear Poisson's equation (i.e., $\mu=1$ in Eq.\ \ref{Eq: Momentum_Carreau}) can be found  analytically (see \citet{Cornish28}), the presence of non-Newtonian viscosity makes Eq. (\ref{Eq: Momentum_Carreau}) strongly non-linear and the problem can only be solved numerically.

%%%%%%%%%%%%%%%%%%%%%%%%%%%%%%%%%%%%%%%%%%%%%%%%%%%%%%%%%%%%%%%%%%%%%%%%%%%%%%%%%%%%%%%%%%%%
\section{ Numerical method}\label{Sec: Numerics}

In order to solve numerically the strongly non-linear problem, the computations are divided into two  nested iteration loops. In the inner iterations, Eq.\ (\ref{Eq: Momentum_Carreau}) is solved for the velocity profile $u(x,y)$ with a given value of the pressure gradient $dp/dz$, while in the outer iteration loop $dp/dz$ is corrected to satisfy the flow rate constraint (Eq. \ref{Eq: Flow_rate_constraint}).

The problem is solved numerically using the finite-volume discretization in the rectangular flow domain of dimensions $\displaystyle[0..1/\varepsilon,0..1]$. The velocity is calculated at the center of each rectangular cell, $\displaystyle[x_i,y_j]$, of size $\displaystyle\Delta x_i$ and $\displaystyle\Delta y_j$, where $\displaystyle i=0,...,N_x$ and $\displaystyle j=0,...,N_y$, respectively. To deal with the non-linearity of Eq.\ (\ref{Eq: Momentum_Carreau}), we use the Newton's method to solve the system of $(N_x-1)\times(N_y-1)$ discretized algebraic equations of the form:
\begin{equation} \label{Eq: Numerical}
    \frac{\partial}{\partial x} \bigg(\mu \frac{\partial u}{\partial x}\bigg) \Biggr|_{i,j}
    + \frac{\partial}{\partial y} \bigg(\mu \frac{\partial u}{\partial y}
    \bigg) \Biggr|_{i,j}
    - \frac{d p}{d z} 
    = 0.
\end{equation}
The successive iterations for the velocity (with a particular pressure gradient value) are then obtained by solving
\begin{equation} \label{Eq: Newton}
	\vec{J_F}(\vec{U}^{k}) \bigl(\vec{U}^{k+1}
	- \vec{U}^{k}\bigr)
	= - \vec{F} (\vec{U}^{k}),
\end{equation}
where $\displaystyle\vec{U}^{k}$ is a matrix-form notation of the discretized base flow velocity at iteration $k$, $\displaystyle u_{i,j}^{k}$, $\displaystyle\vec{F}$ is a vector-form notation for the left-hand-side of Eq.\ \ref{Eq: Numerical}, and its Jacobian, $\vec{J_F}$, with respect to the velocity vector is defined as
\begin{equation} \label{Eq: Jacobian}
	J_F(\vec{U}^{k}) = \frac{\partial \vec{F} (\vec{U}^{k})}{\partial \vec{U}^{k}}.
\end{equation}
As a result of the second-order central finite volume discretization, the expression for $\displaystyle\vec{F}$ at iteration $k$ reads
\begin{align}
	&\begin{aligned}
		F_{i,j}^{k} 
		&= m 
		\biggl(\frac{u_{i+1,j}^{k} - 2 u_{i,j}^{k} + u_{i-1,j}^{k}}{\Delta x_i^2}
		+ \frac{u_{i,j+1}^{k} - 2 u_{i,j}^{k} + u_{i,j-1}^{k}}{\Delta y_j^2}\biggr)
		\\
		&+ \frac{1-m}{\Delta x_i^2} 
		\Biggl(1+ \Cu^2 
		\biggl[\bigg(\frac{u_{i+1,j}^{k}-u_{i,j}^{k}}{\Delta x_i}\bigg)^2
		+ \bigg(\frac{u_{i,j+1}^{k}-u_{i,j-1}^{k}}{2 \Delta y_j}\bigg)^2\biggr] \Biggr)^{(n-1)/2}
		\biggl(u_{i+1,j}^{k}-u_{i,j}^{k}\biggr)
		\\
		&- \frac{1-m}{\Delta x_i^2}
		\Biggl(1+ \Cu^2 
		\biggl[\bigg(\frac{u_{i,j}^{k}-u_{i-1,j}^{k}}{\Delta x_i}\bigg)^2
		+ \bigg(\frac{u_{i,j+1}^{k}-u_{i,j-1}^{k}}{2 \Delta y_j}\bigg)^2\biggr] \Biggr)^{(n-1)/2}
		\biggl(u_{i,j}^{k}-u_{i-1,j}^{k}\biggr)
		\\
		&+ \frac{1-m}{\Delta y_j^2} 
		\Biggl(1+ \Cu^2 
		\biggl[\bigg(\frac{u_{i+1,j}^{k}-u_{i-1,j}^{k}}{2 \Delta x_i}\bigg)^2
		+ \bigg(\frac{u_{i,j+1}^{k}-u_{i,j}^{k}}{\Delta y_j}\bigg)^2\biggr]
		\Biggr)^{(n-1)/2}
		\biggl(u_{i,j+1}^{k}-u_{i,j}^{k}\biggr)
		\\
		&- \frac{1-m}{\Delta y_j^2}
		\Biggl(1+ \Cu^2 
		\biggl[\bigg(\frac{u_{i+1,j}^{k}-u_{i-1,j}^{k}}{2 \Delta x_i}\bigg)^2
		+ \bigg(\frac{u_{i,j}^{k}-u_{i,j-1}^{k}}{\Delta y_j}\bigg)^2\biggr] \Biggr)^{(n-1)/2}
		\biggl(u_{i,j}^{k}-u_{i,j-1}^{k}\biggr)	
		\\
		&- \frac{ dp}{ dz}^{k} \, .
	\end{aligned}
\end{align}
Then the Jacobian matrix (Eq.\ \ref{Eq: Jacobian}) is a sparse square matrix of size $N_x+N_y+N_x N_y$. For the chosen discretization scheme, the Jacobian matrix has a banded five-diagonal structure. The corresponding system of linear algebraic equations is solved by a direct solver for sparse linear equation systems (MUMPS package in FORTRAN) at each iteration of Newton's method \cite{Gelfgat07a}. As the initial guess for the Newton's method ($k=0$), we use the velocity profile and the corresponding pressure gradient of a Newtonian fluid, $\displaystyle(-dp/dz)_N$, given by \citet{Cornish28}.

Once the inner problem is solved for a given pressure gradient, the results are fed into 
the outer iteration loop that applies the secant method to correct the pressure gradient, until it converges to the proper value for which the corresponding velocity field satisfies the flow rate constraint. The convergence criterion for the pressure gradient is set for the relative difference between iterations of the outer loop to be less than $10^{-5}$, while the convergence of the velocity field is achieved when its absolute change from iteration to iteration of the inner problem becomes less than $10^{-10}$. In the following, we use a computational grid with $N_x = N_y = 100$ and $\sin$-stretching near the walls, $\displaystyle x \rightarrow 0.06\sin(2\pi x)$ and $\displaystyle y \rightarrow 0.06\sin(2\pi y)$, so that the computational cells of the smallest size, $\min(\Delta y_j)\approx0.003$ (compare with $\Delta y_j = 0.01$ for the uniform grid), lie in the boundary layer. The largest cells are then located around the horizontal centerline and their vertical size does not exceed $0.014$. This allows obtaining at least three correct digits after the decimal point in the pressure gradient.

%%%%%%%%%%%%%%%%%%%%%%%%%%%%%%%%%%%%%%%%%%%%%%%%%%%%%%%%%%%%%
\section{Results and discussion}\label{Sec: Results} 

In a rectangular duct, the steady-state solution for the fully-developed laminar flow of shear-thinning fluid  depends on its rheology and the duct aspect ratio. In the following, we discuss how these two factors affect both the axial velocity field in the duct cross-section and integral flow characteristics, such as the pressure gradient and friction factor.

%%%%%%%%%%%%%%%%%%%%%%%%%%%%%%%%%%%%%%%%%%%%%%%%%%%%%%%%%%%%%%%%%%%%%%%%%%%
\subsection{Effect of the shear-thinning rheology}\label{Sec: Results_a}

In this section, we consider the effect of the shear-thinning rheology on the flow between two plates, which is a particular case of a rectangular duct with very small height-to-width aspect ratio, i.e., $\displaystyle\varepsilon=H/W\to0$. The study of this limiting case is convenient since, as it will be shown below (Section\ \ref{Sec: Results_b}), the effect of rheology is independent of the particular aspect ratio considered.

\begin{figure}[]
	\centering
	\subfloat[Pressure gradient]{\includegraphics[width=0.5\textwidth,clip]{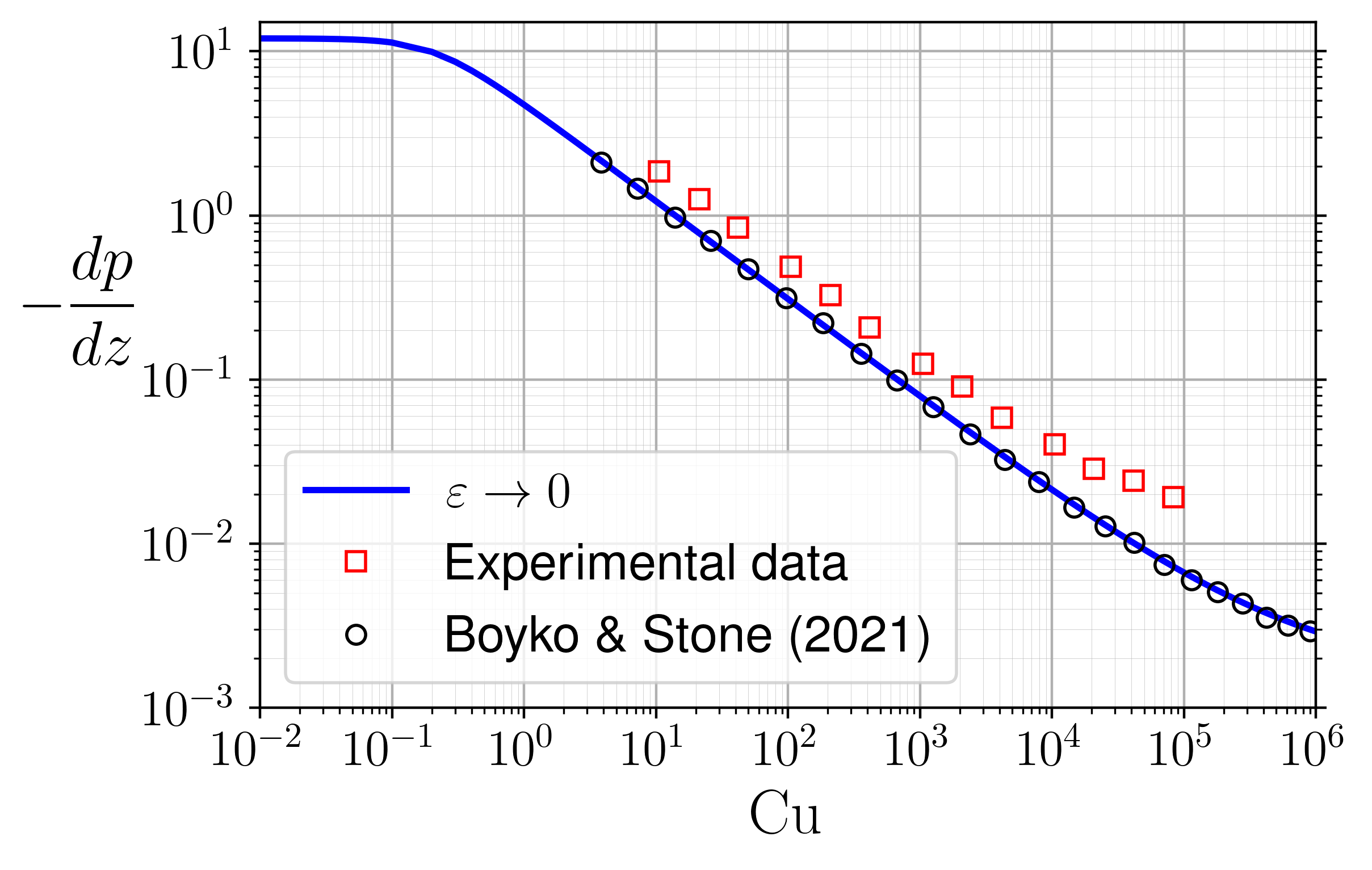}}
	\subfloat[Velocity]{\includegraphics[width=0.48\textwidth,clip]{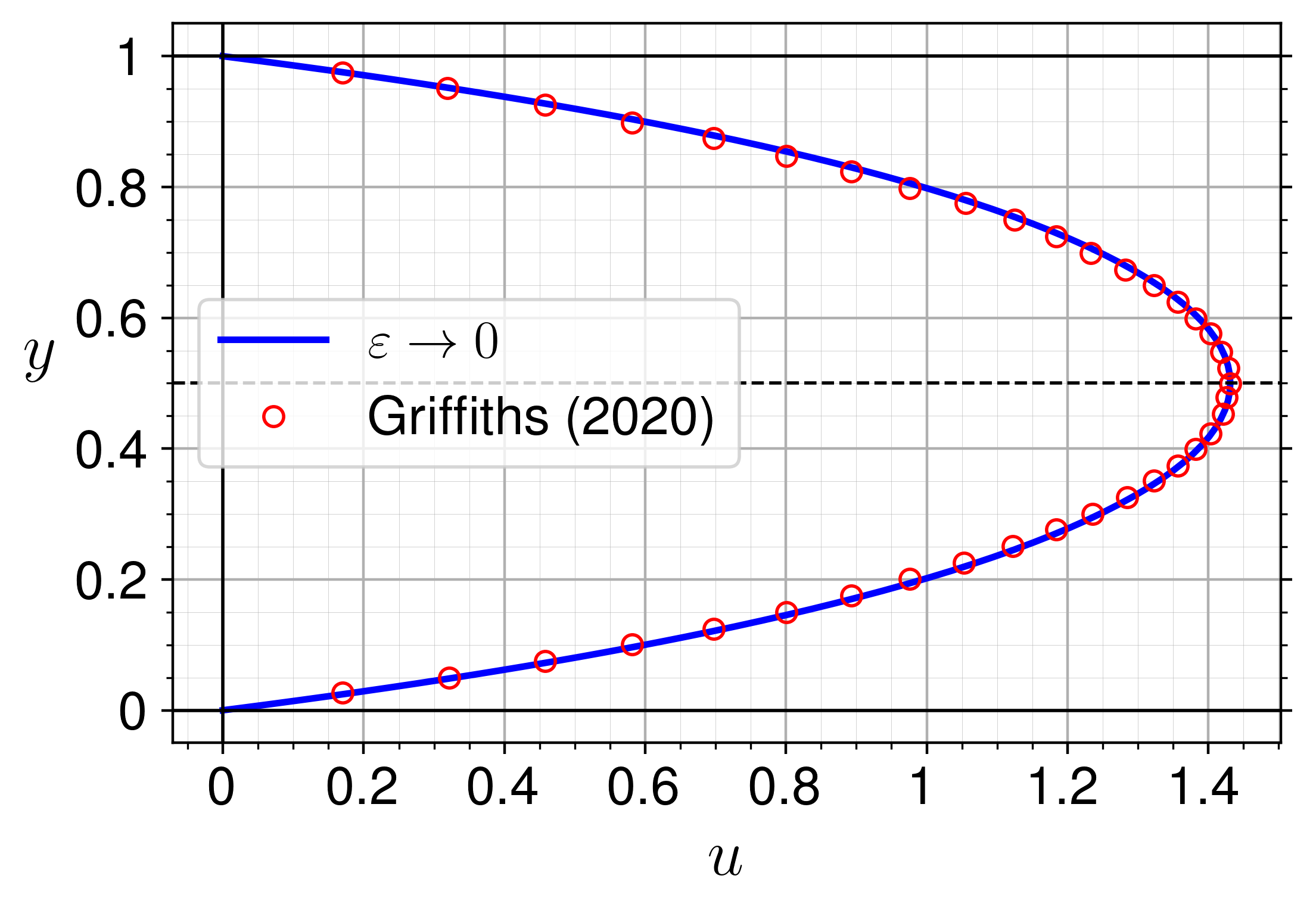}}
	\caption{\label{Fig: Validation}(a) Dimensionless pressure gradient as a function of Carreau number. Comparison of the experimental data from \citet{Pipe08} and theoretical results reported in \citet{Boyko21} with the present numerical results for the two-plate limit ($\varepsilon\to0$). Rheological parameters of the shear-thinning liquid: $m=0.000135, n=0.402$. (b) Comparison of the velocity profile obtained in the present study for $\varepsilon\to0$ with that presented in \citet{Griffiths20} in Fig. 2a. Rheological parameters: $m=0, n=1/3$. In this work $\lambda=1$ of \cite{Griffiths20} corresponds to $\Cu=0.25$.} 
\end{figure}

The two-plate limit serves also as a validation of our numerical results obtained in the limit of $\displaystyle\varepsilon\to0$. Comparison with the results of \citet{Picchi17} (for the single-phase flow) gives an excellent agreement for the velocity profile and the pressure gradient (the validation is not shown here for the sake of brevity). In Fig.\ \ref{Fig: Validation}a we show the comparison between our numerical predictions for the pressure gradient (normalized with the scaling used in this study), the experimental data of \citet{Pipe08} collected in a microchannel of aspect ratio $\varepsilon\approx0.008$ (the test fluid is an aqueous xanthan gum solution, where $m=0.000135$ and $n=0.402$, see \cite{Pipe08,Boyko21}), and the theoretical predictions of \citet{Boyko21} for the two-plate geometry. The latter overlaps with the present numerical results, which, however, underestimates the results of the experiment of \citet{Pipe08}, as was already discussed in detail in \citet{Boyko21}. Note that the pressure gradient is negative along the duct axis, i.e., $dp/dz<0$, and the pressure gradient values presented are those of $(-dp/dz)$.

The shear-thinning rheology is determined by the dimensionless parameters $m$ and $n$, and by Carreau number defined in Eq.\ (\ref{Eq: Cu}). Specifically, the Carreau number controls the onset of the shear-thinning effect. When $\Cu \to 0$,  the fluid rheology is dominated by the zero-shear-rate viscosity ($\mu=1$) and the dimensionless pressure gradient is the same as of a Newtonian fluid, i.e., $\displaystyle -dp/dz=12$, as can be seen in Fig.\ \ref{Fig: Validation}a. At intermediate $\Cu$, the pressure gradient decreases with $\Cu$ in a similar manner as the viscosity in the power-law region, while, for $\Cu \to \infty$ the infinity shear-rate viscosity dominates and pressure gradient flattens to a constant value of  $-dp/dz=12 m$. In this limit, the Newtonian pressure gradient of a fluid with the infinite-shear-rate viscosity ($\mu=m$) is recovered. 

In the two-plate limit, we can also validate the velocity profile by comparison to the analytical solution presented by \citet{Griffiths20}, that was obtained for $\displaystyle m=0$ ($\displaystyle\mu_\infty\ll\mu_0$) and a discrete set of values of $n$, such that $\displaystyle 6/(1-n)\in\mathbb{N}$. In Fig. \ref{Fig: Validation}b, the numerical (for $\displaystyle\varepsilon\to0$) and analytical velocity distributions are shown for $\displaystyle\Cu=0.25$, which corresponds to the case of $\displaystyle\lambda=1$, presented in the dimensional form in Fig. 2a in \cite{Griffiths20}. Such an agreement between the numerical and analytical results is obtained also for the other values of $\lambda$ presented in that figure (rendered dimensionless using the scaling of the present work, i.e., $\displaystyle\lambda=0.5$ and $1.5$ correspond to $\displaystyle\Cu=0.125$ and $0.375$, respectively). The corresponding dimensionless velocity profiles are close to each other and not shown in Fig.\ \ref{Fig: Validation}b.

\begin{figure}[]
    \centering    
    \subfloat[Velocity]{\includegraphics[width=0.48\textwidth,clip]{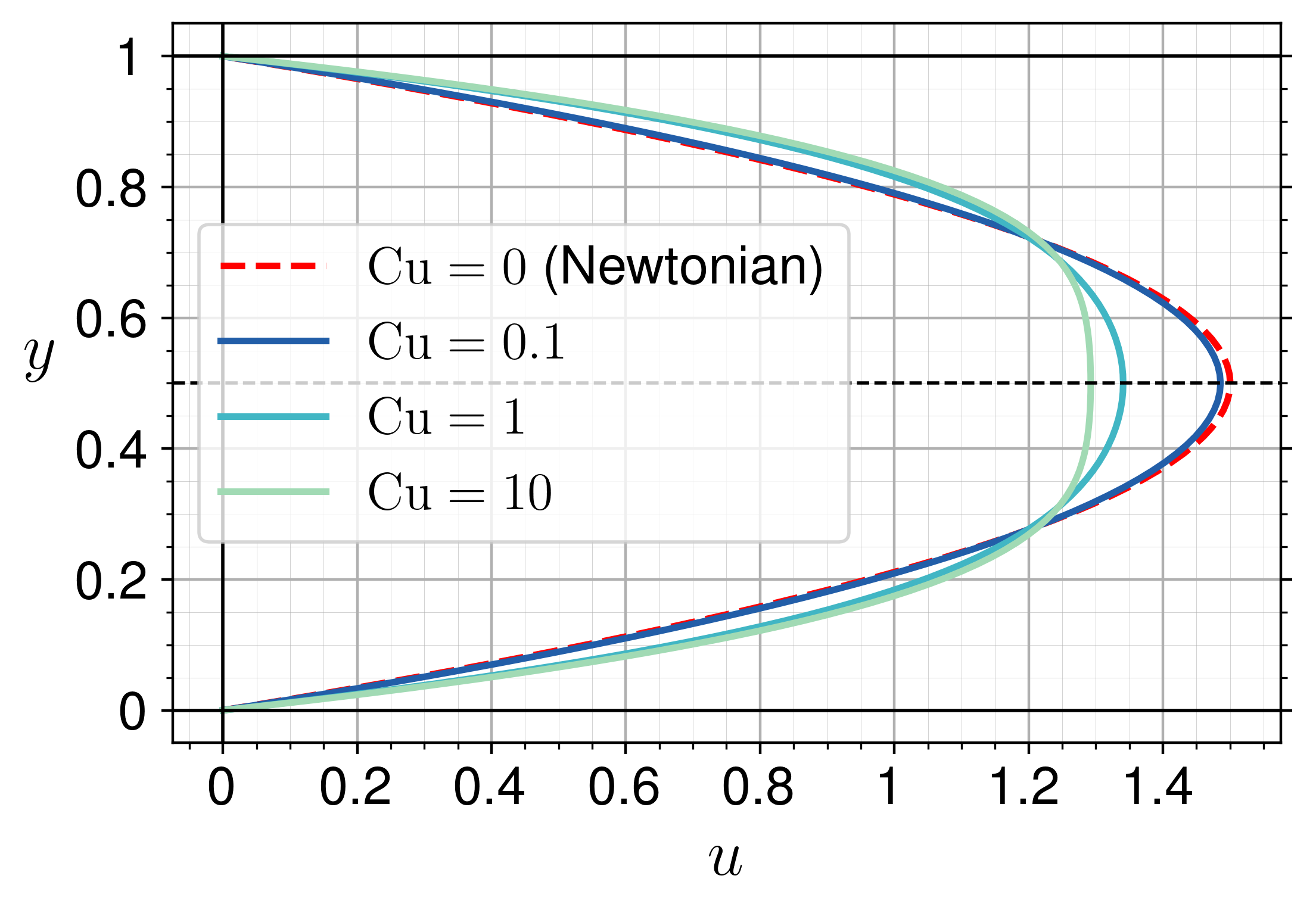}}
    \subfloat[Viscosity]{\includegraphics[width=0.48\textwidth,clip]{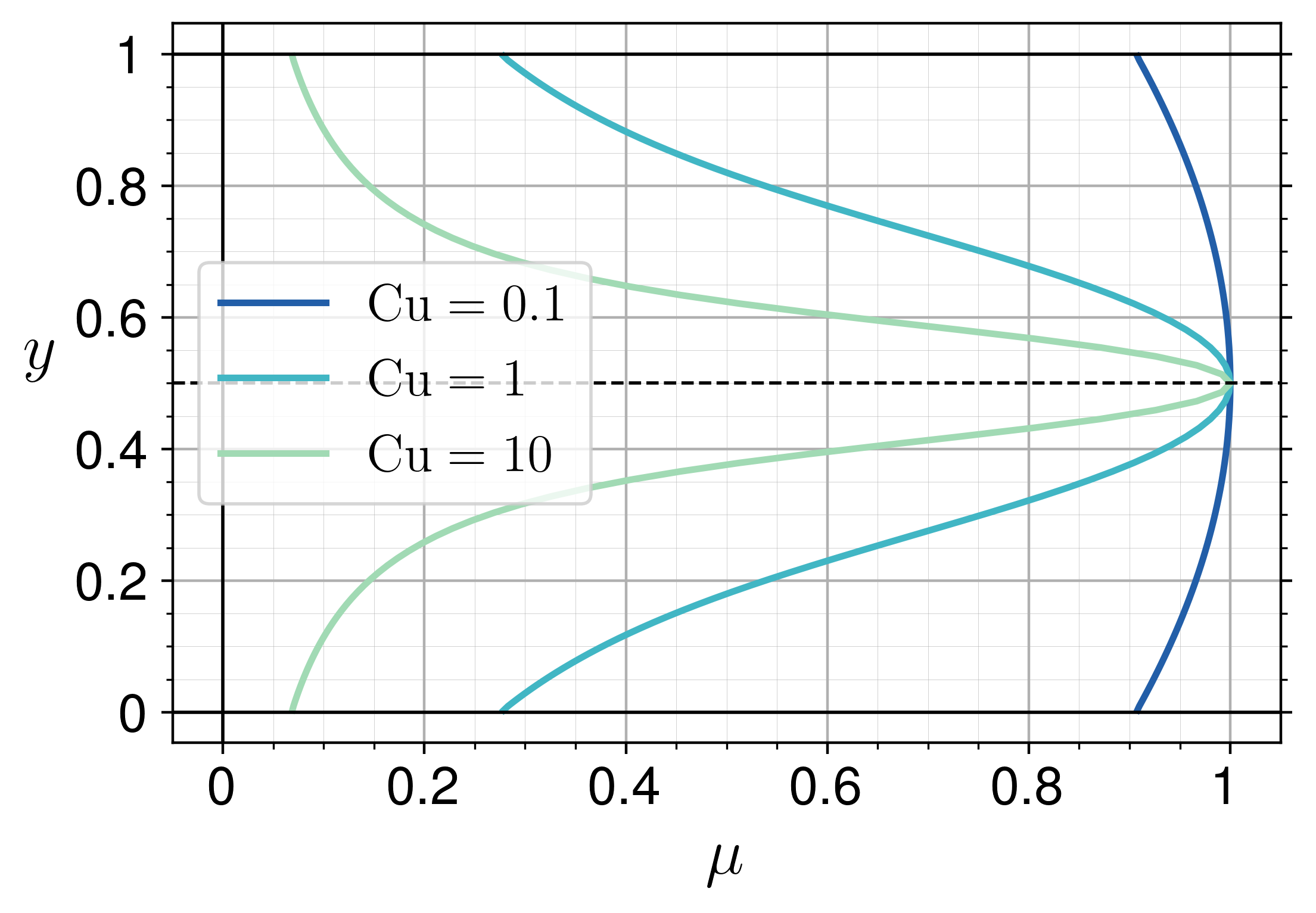}}
	\caption{\label{Fig: TP_profiles}Profiles of the velocity (a) and viscosity (b) on the vertical centerline of xanthan gum solution ($m = 0.000135$, $n=0.402$) flowing in a rectangular duct in the two-plate geometry limit ($\displaystyle\varepsilon\to0$). Effect of Carreau number.} 
\end{figure}

The velocity and viscosity profiles in the limit of $\varepsilon\to0$  are shown in Fig.\ \ref{Fig: TP_profiles} for different values of Carreau number. As expected, for $\Cu=0$, the parabolic Newtonian profile reaches the maximum velocity at $u_{\max}=1.5$, while larger values of $\Cu$ result in flattening of the velocity distribution in the channel center. For $\Cu=0.1$, which corresponds to the beginning of the transitional region between the  Newtonian and the power-law behaviors (see Fig.\ \ref{Fig: Validation}), the deviation from the Newtonian value is rather subtle, $u_{\max}\approx1.48$, while it is getting larger for $\Cu=1$ and $10$ ($u_{\max}\approx1.34$ and $u_{\max}\approx1.29$, respectively). Recovery of the parabolic velocity profiles is obtained at much higher $\Cu$ (not shown). The impact of the shear-thinning effect is clearer when examining how far the viscosity $\mu$ deviates from the two Newtonian limits (Fig.\ \ref{Fig: TP_profiles}b). Due to the flow symmetry, the shear rate is always zero at the center, where the fluid exhibits the corresponding Newtonian viscosity ($m=1$). The shear rate increases towards the walls, where stronger shear-thinning effects are observed.

\begin{figure}[]
    \centering    
    \subfloat[$n=0.402$]{\includegraphics[width=0.5\textwidth,clip]{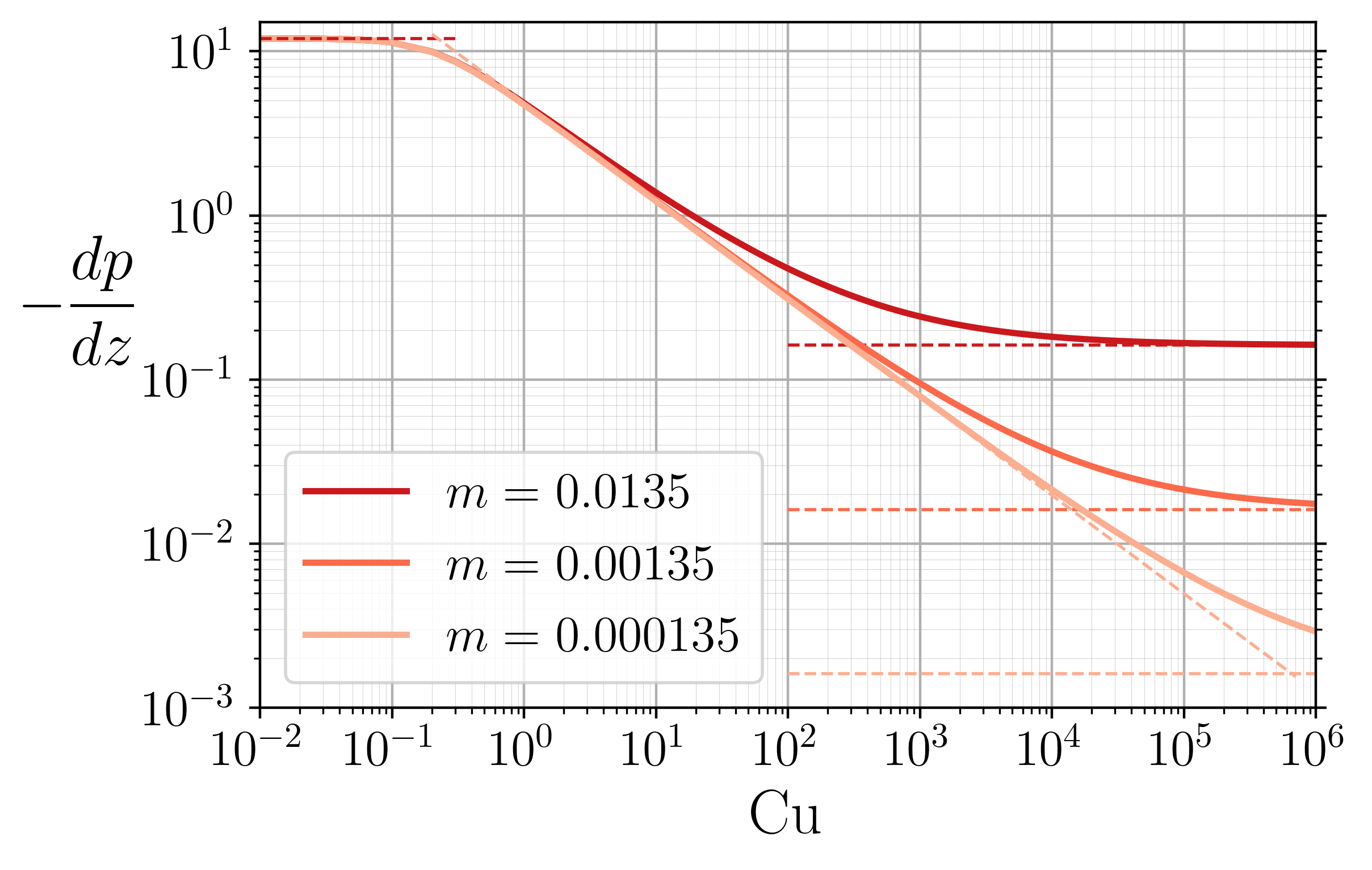}}
    \subfloat[$m=0.000135$]{\includegraphics[width=0.5\textwidth,clip]{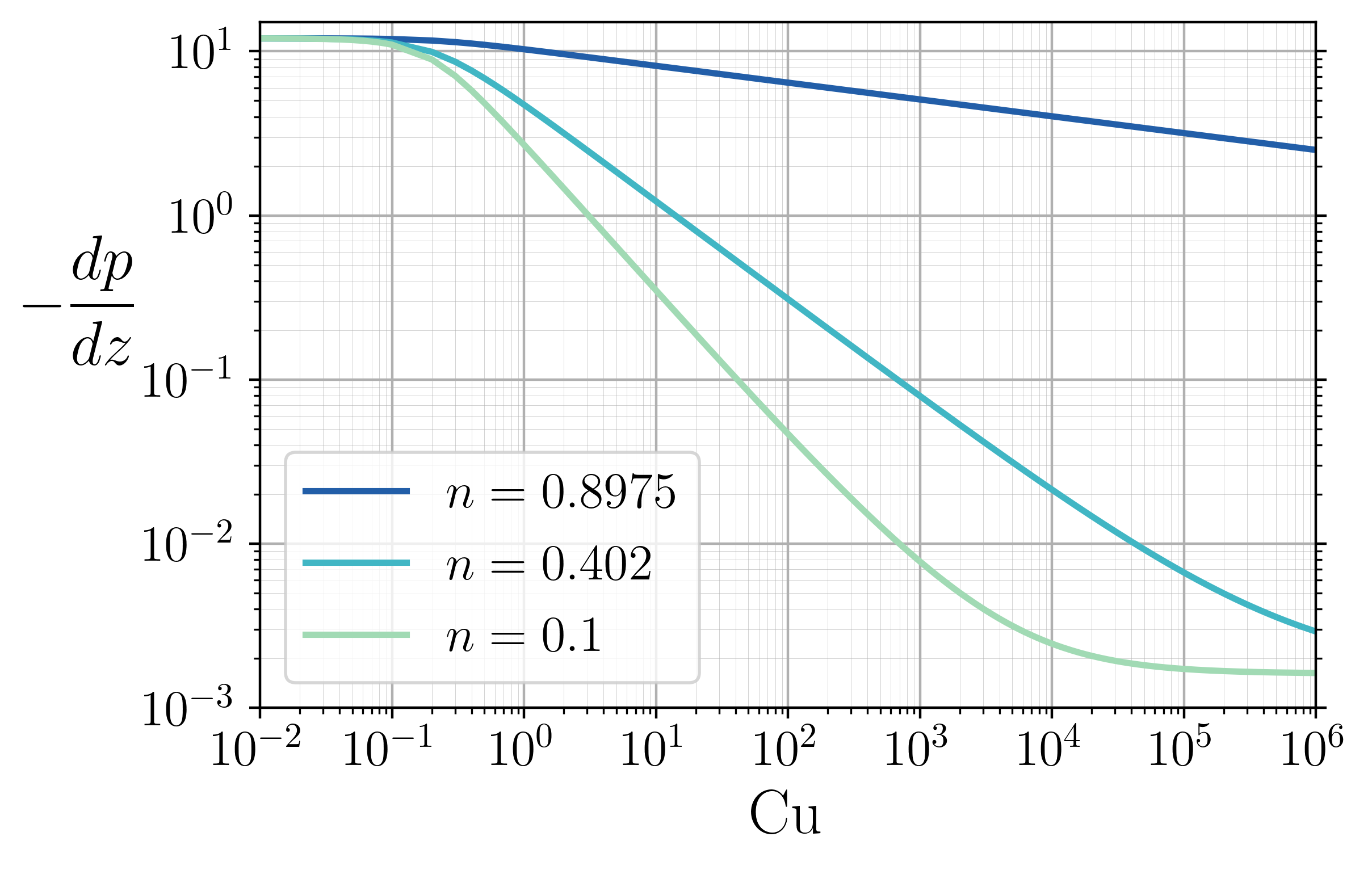}}
    \caption{ \label{Fig: Rheology}Dimensionless pressure gradient as a function of Carreau number for a rectangular duct in the two-plate limit ($\varepsilon \to 0$). (a) Effect of different ratios of the zero- to infinite-shear-rate viscosity, $m$; (b) effect of the shear-thinning index, $n$.}
\end{figure}

The variation of the velocity and viscosity profiles with the Carreau number has an overall effect on the pressure gradient, $(-dp/dz)$, which is  a function of the Carreau number, the shear-thinning index, and the aspect ratio only, i.e.,  $dp/dz=f({\rm Cu},n,m,\varepsilon)$. As shown in Fig.\ \ref{Fig: Rheology}a, the dependence of $-dp/dz$ on $\Cu$ (on log-log plot) follows the rheological equation of the shear-thinning liquid (Eq.\ \ref{Eq: Mu}). The Newtonian (for $\mu=1$ and $\mu=m$) and the power-law asymptotes (with a slope of $n-1$ in logarithmic coordinates) are shown in the figure as dashed lines. The value of the parameter $m$ affects the width of the power-law region: 
for smaller $m$ and fixed $n$ (e.g., $n=0.402$), the shear-thinning behavior prevails over a wider range of Carreau numbers. The shear-thinning index $n$, on the other hand, impacts the slope of the viscosity versus $\Cu$ curve, and, as a result, the slope of the pressure gradient curve in the power-law region. When $m$ is fixed (e.g., $m=0.000135$ in Fig. \ref{Fig: Rheology}(b)), the slope increases with decreasing $n$, resulting in shrinkage of the power-law region and a shift of the high-shear-rate Newtonian plateau to smaller $\Cu$. 

%%%%%%%%%%%%%%%%%%%%%%%%%%%%%%%%%%%%%%%%%%%%%%%%%%%%%%%%%%%%%%%%%%%%%%%%%%%
\subsection{\label{Sec: Results_b}Effect of the aspect ratio}

In order to examine the effect of the aspect ratio, $\varepsilon$, on the flow characteristics, we refer here to a particular Carreau fluid (xanthan gum solution), with the rheological parameter values set to $n=0.402$ and $m=0.000135$. Due to the symmetry of the problem, the $x$ and $y$-coordinates are completely interchangeable, and, hence our analysis spans aspect ratio in the range of $\varepsilon=H/W\in(0,1]$, where $\varepsilon\to0$ and $\varepsilon=1$ correspond to the two-plate geometry and square duct, respectively. In the following, we will refer to three representative cases: $\varepsilon\to0$ and $\varepsilon=0.5,\, 1$. 

\begin{figure}[]
    \centering    
    \subfloat[$u$, $\Cu=0.1$]{\includegraphics[width=0.33\textwidth,clip]{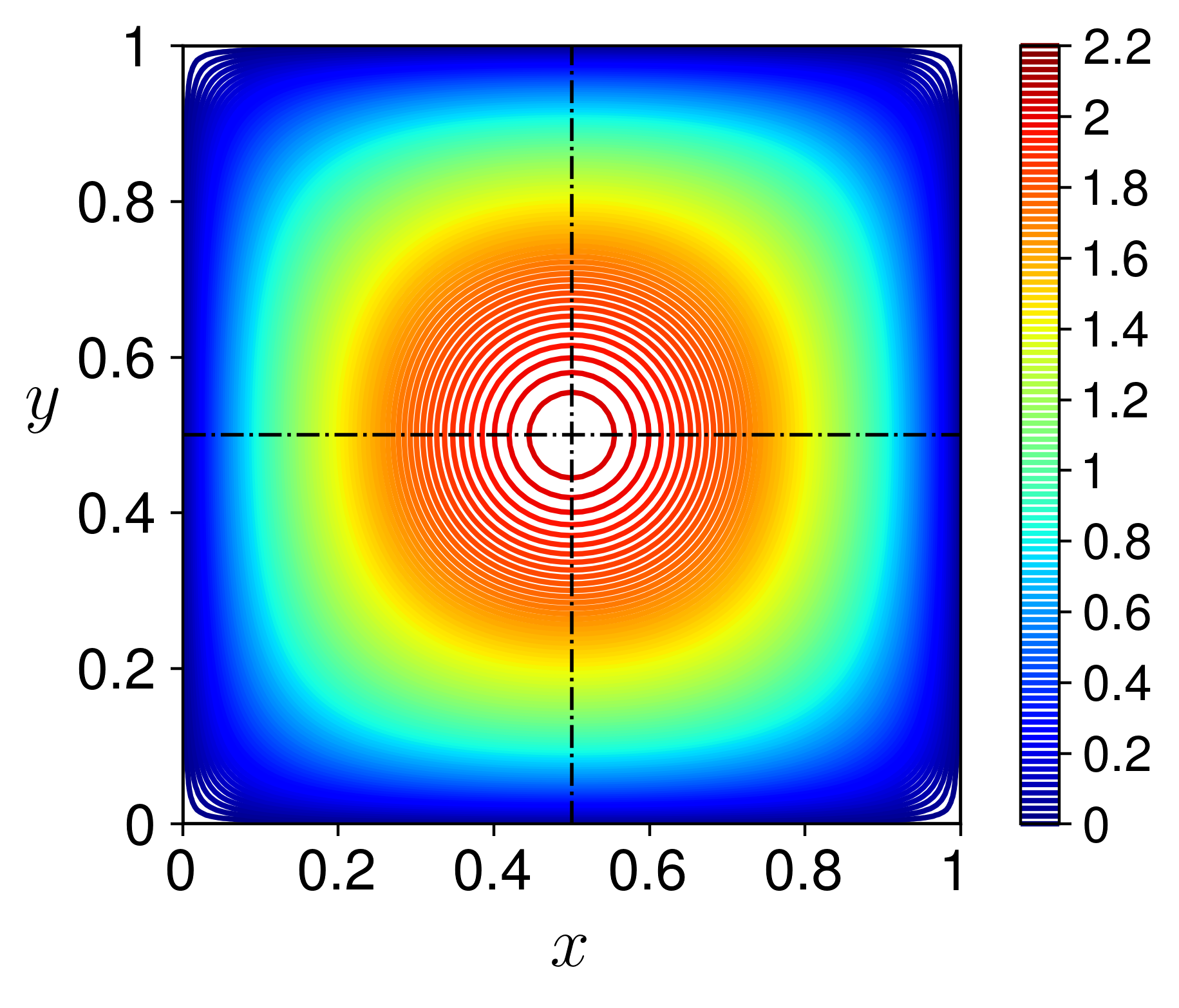}}
    \subfloat[$u$, $\Cu=1$]{\includegraphics[width=0.33\textwidth,clip]{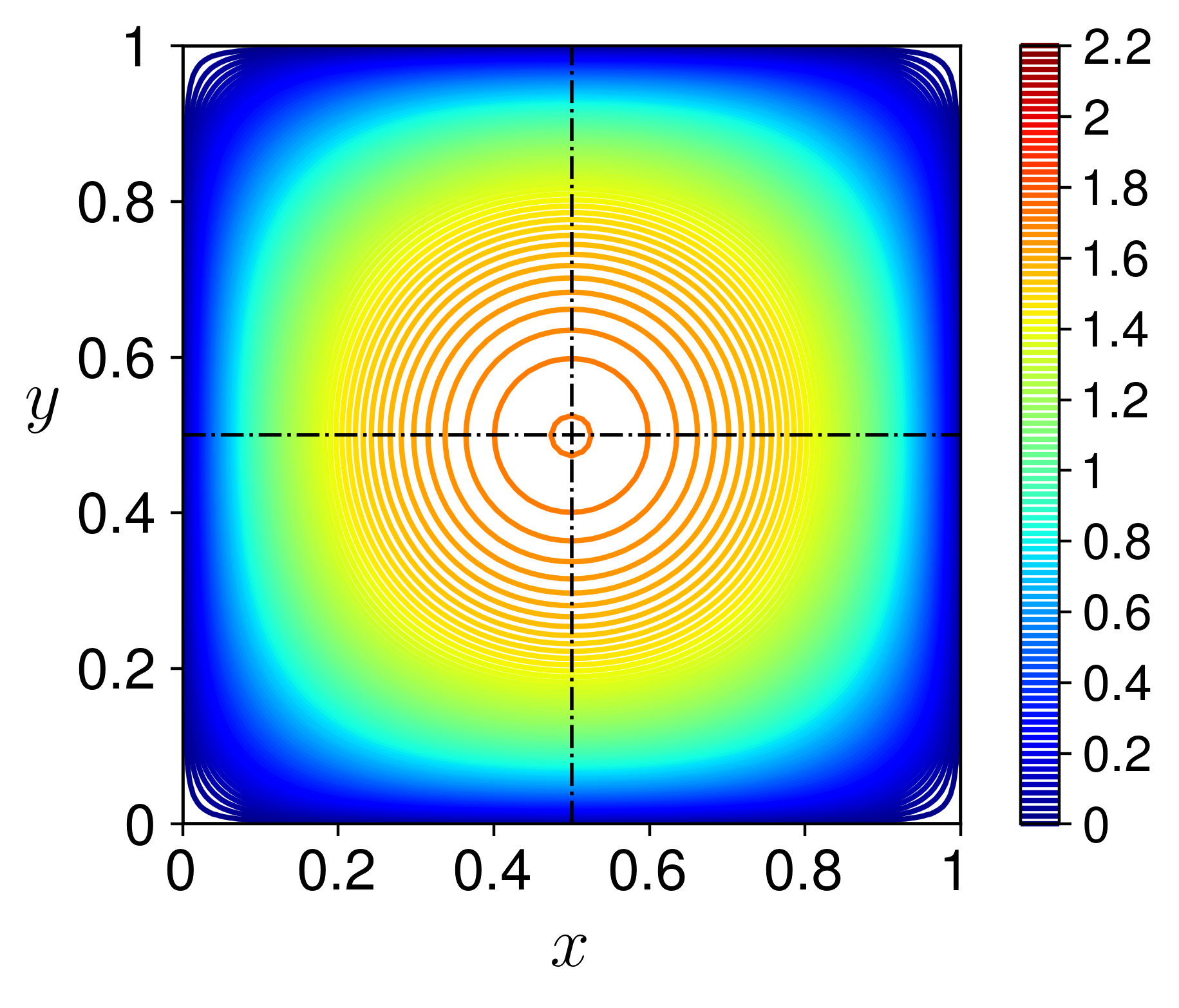}}
    \subfloat[$u$, $\Cu=10$]{\includegraphics[width=0.33\textwidth,clip]{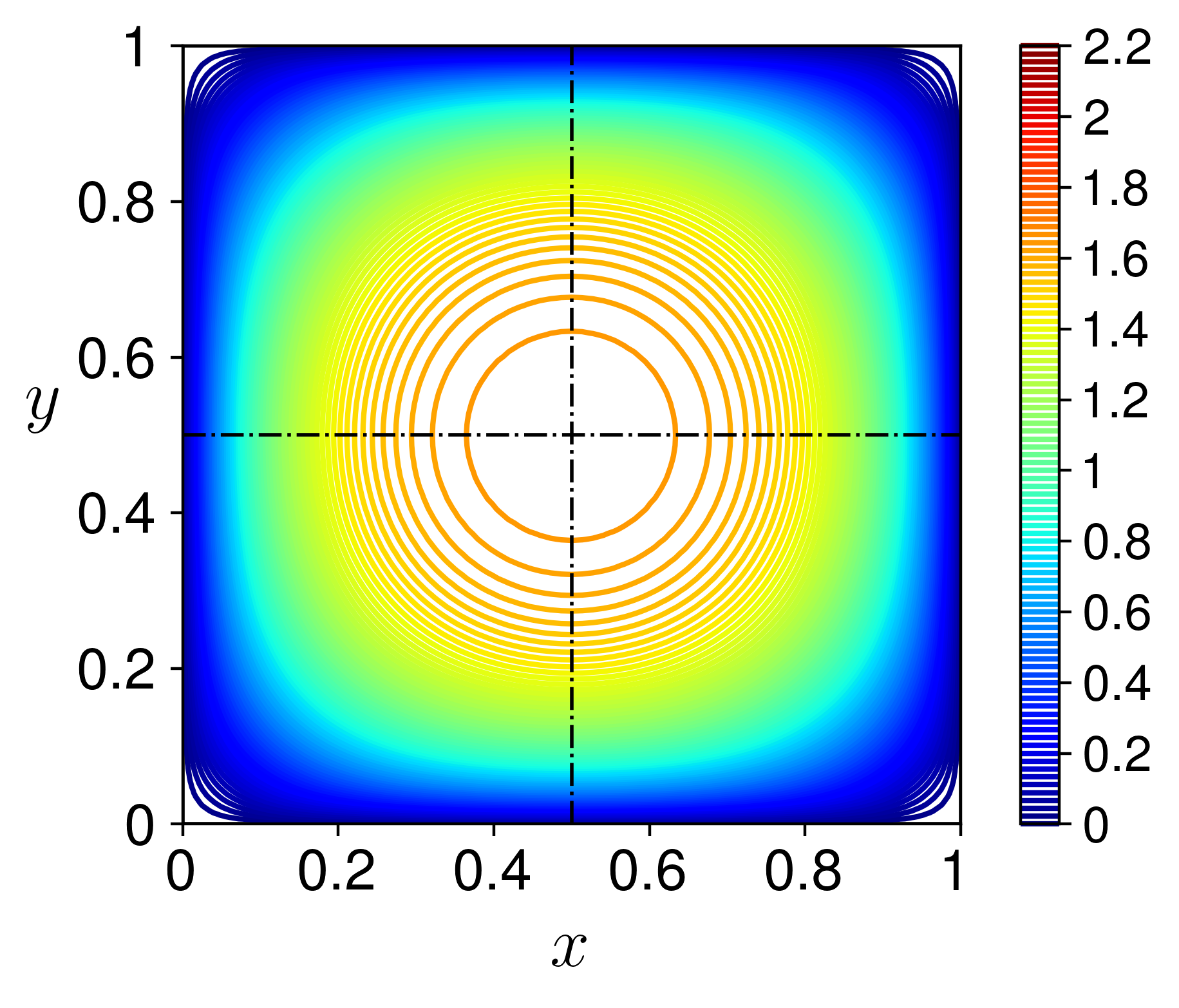}}
    \\
    \subfloat[$\mu$, $\Cu=0.1$]{\includegraphics[width=0.33\textwidth,clip]{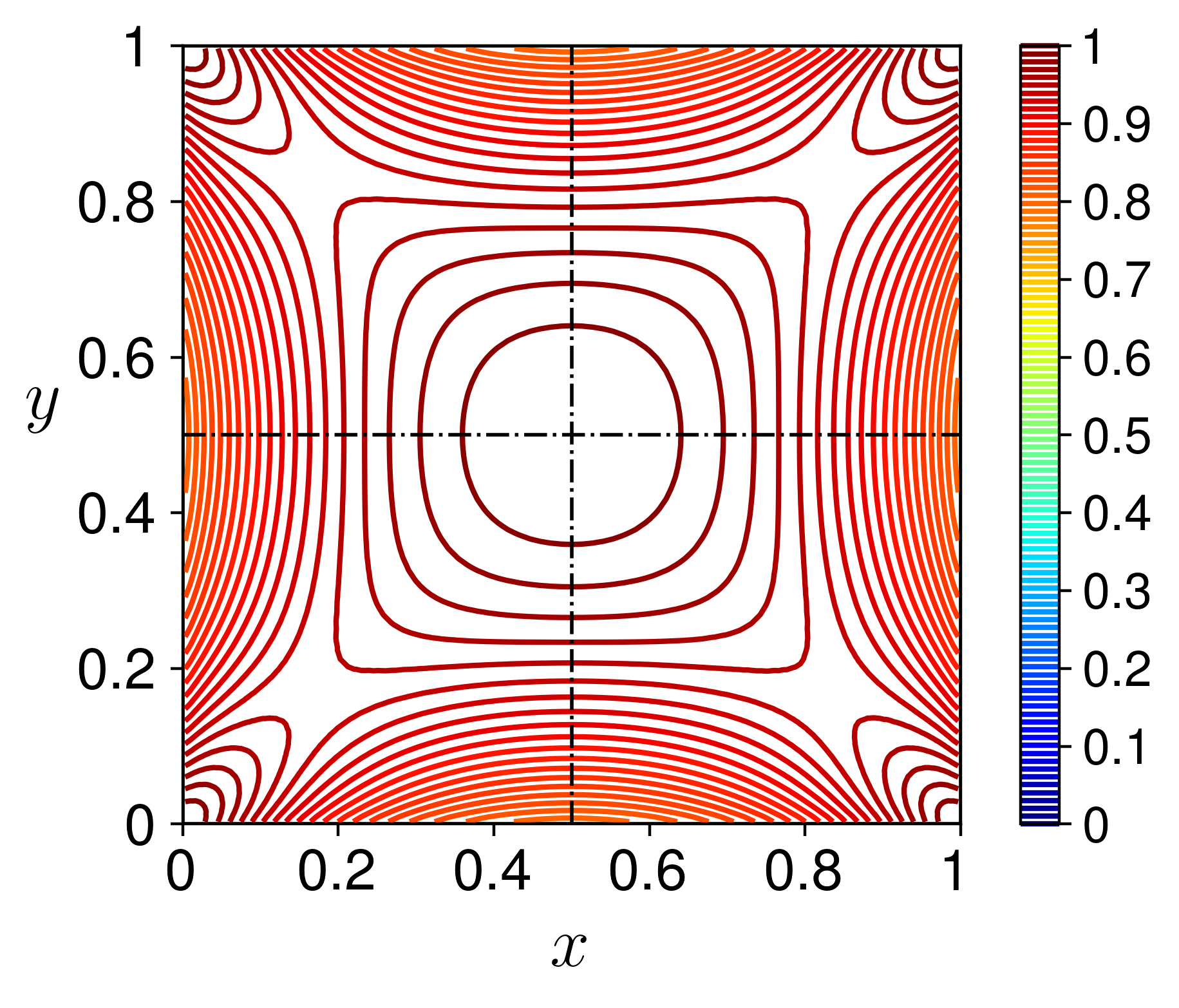}}
    \subfloat[$\mu$, $\Cu=1$]{\includegraphics[width=0.33\textwidth,clip]{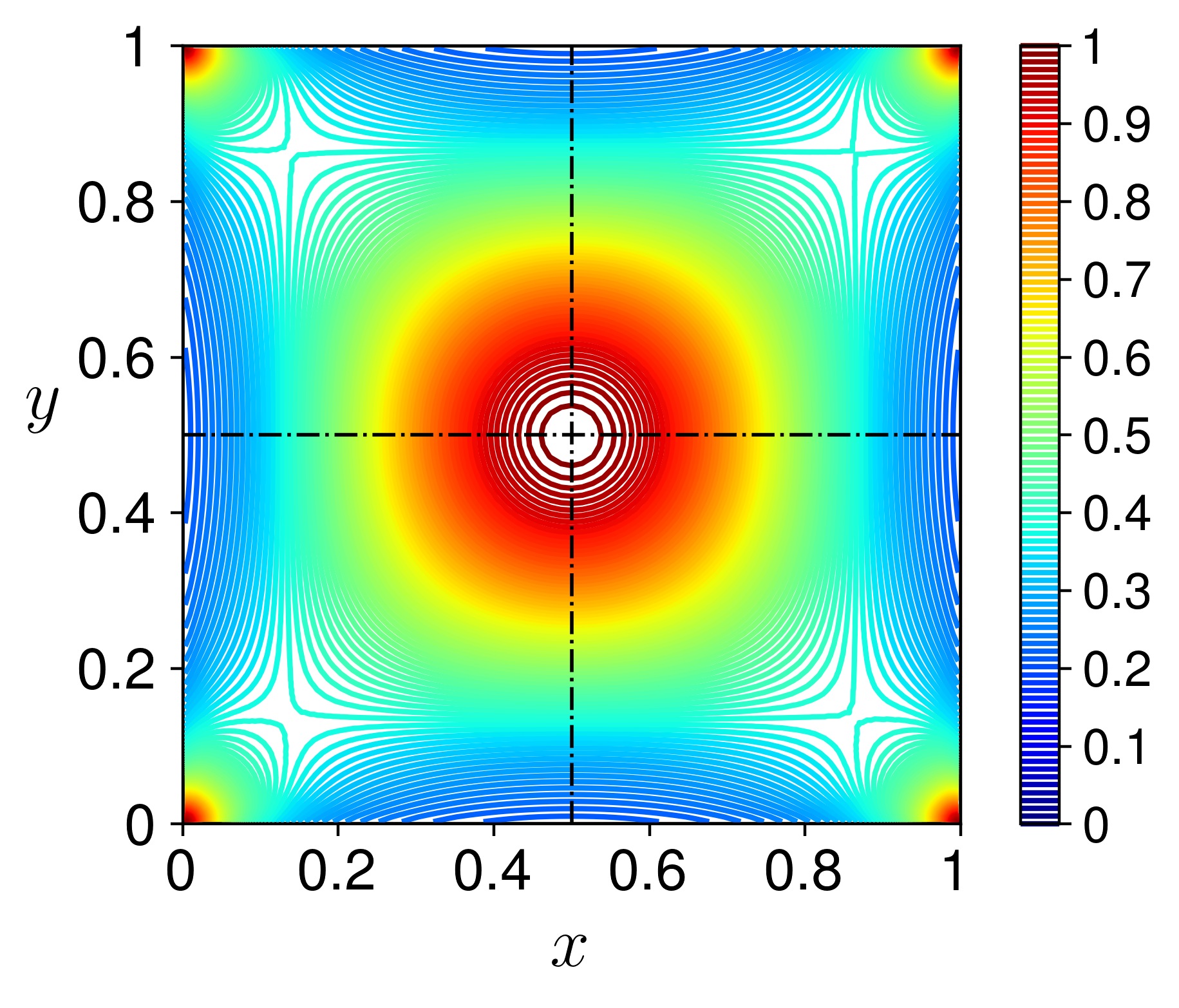}}
    \subfloat[$\mu$, $\Cu=10$]{\includegraphics[width=0.33\textwidth,clip]{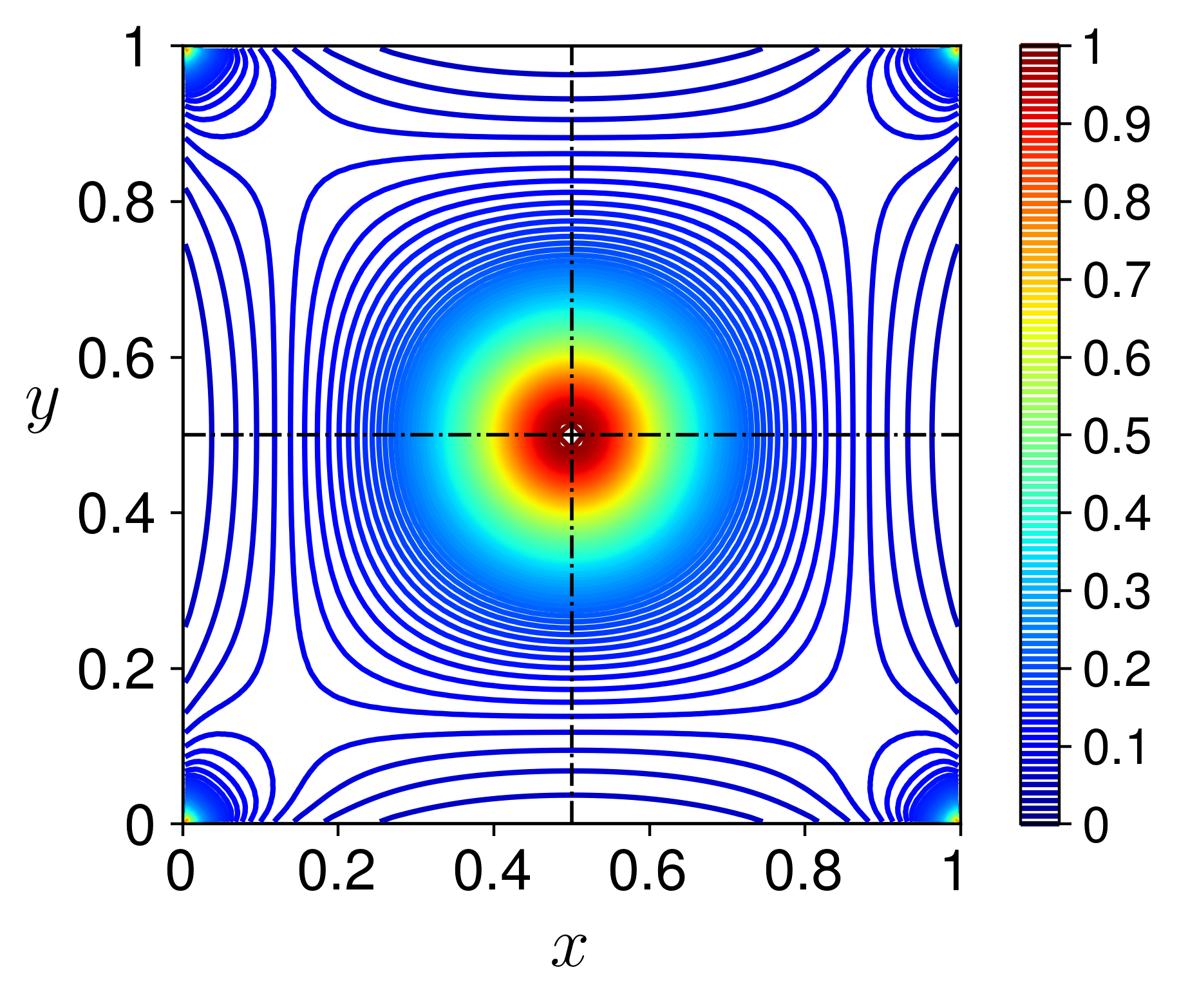}}
    \caption{\label{Fig: Contours}Contours of the velocity (a)-(c) and viscosity (d)-(f) of xanthan gum solution ($m = 0.000135$, $n=0.402$) in a square duct. Effect of Carreau number.} 
\end{figure}

Unlike the two-plate geometry, where the velocity profile is one-dimensional, a finite aspect ratio implies that the velocity field depends on the two cross-sectional coordinates, i.e., $u=u(x,y)$. The numerical solution for a square duct for three different $\Cu$ is shown by color contours in Fig.\ \ref{Fig: Contours}a-c. Obviously, the velocity reaches its maximum in the center of the square. For $\Cu=0.1$, the maximum is equal to $u_{\max}\approx2.04$ and the velocity distribution is slightly different from the Newtonian one, for which $u_{\max}\approx2.1$. The increase in $\Cu$ results in decrease of the maximal velocity (e.g., $u_{\max}\approx1.73$ for $\Cu=1$ and $u_{\max}\approx1.66$ for $\Cu=10$) and flattening of the profile in the center (cf. \ref{Fig: Contours}b,c).

The shape of the velocity profile is related to the cross-sectional distribution of the effective viscosity. The viscosity is almost Newtonian, $\mu=1$, in the whole cross section for $\Cu=0.1$ (Fig.\ \ref{Fig: Contours}d), while it gets smaller (i.e., stronger shear-thinning effects) in large parts of the square for the higher $\Cu=1$ (e) and $\Cu=10$ (f). In fact, with the increase of $\Cu$, the viscosity experiences the fastest decrease from its Newtonian value in the center (dark red contours) to its minimal values next to the walls, e.g., almost ten time decrease for $\Cu=1$ (blue contours in Fig.\ \ref{Fig: Contours}b). Away from the center, the contour lines of the viscosity attain a square shape, due to the smaller velocity gradients and the associated slower viscosity decrease along the diagonals. A local minimum of the viscosity is observed there (blank spaces on the diagonals in the contour plots), followed by its increase towards the corner.

\begin{figure}[]
    \centering    
    \subfloat[Velocity]{\includegraphics[width=0.48\textwidth,clip]{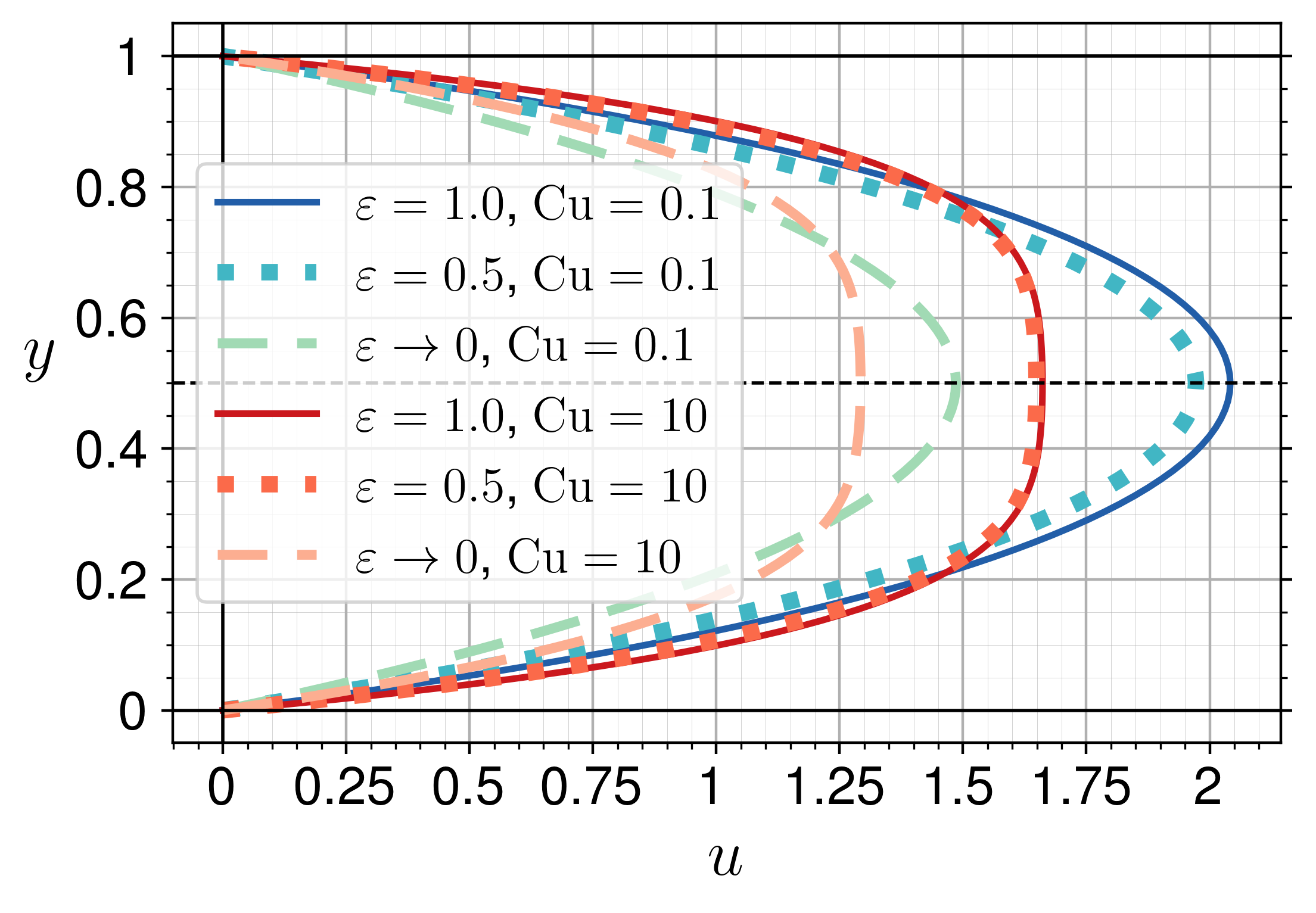}}
    \subfloat[Viscosity]{\includegraphics[width=0.48\textwidth,clip]{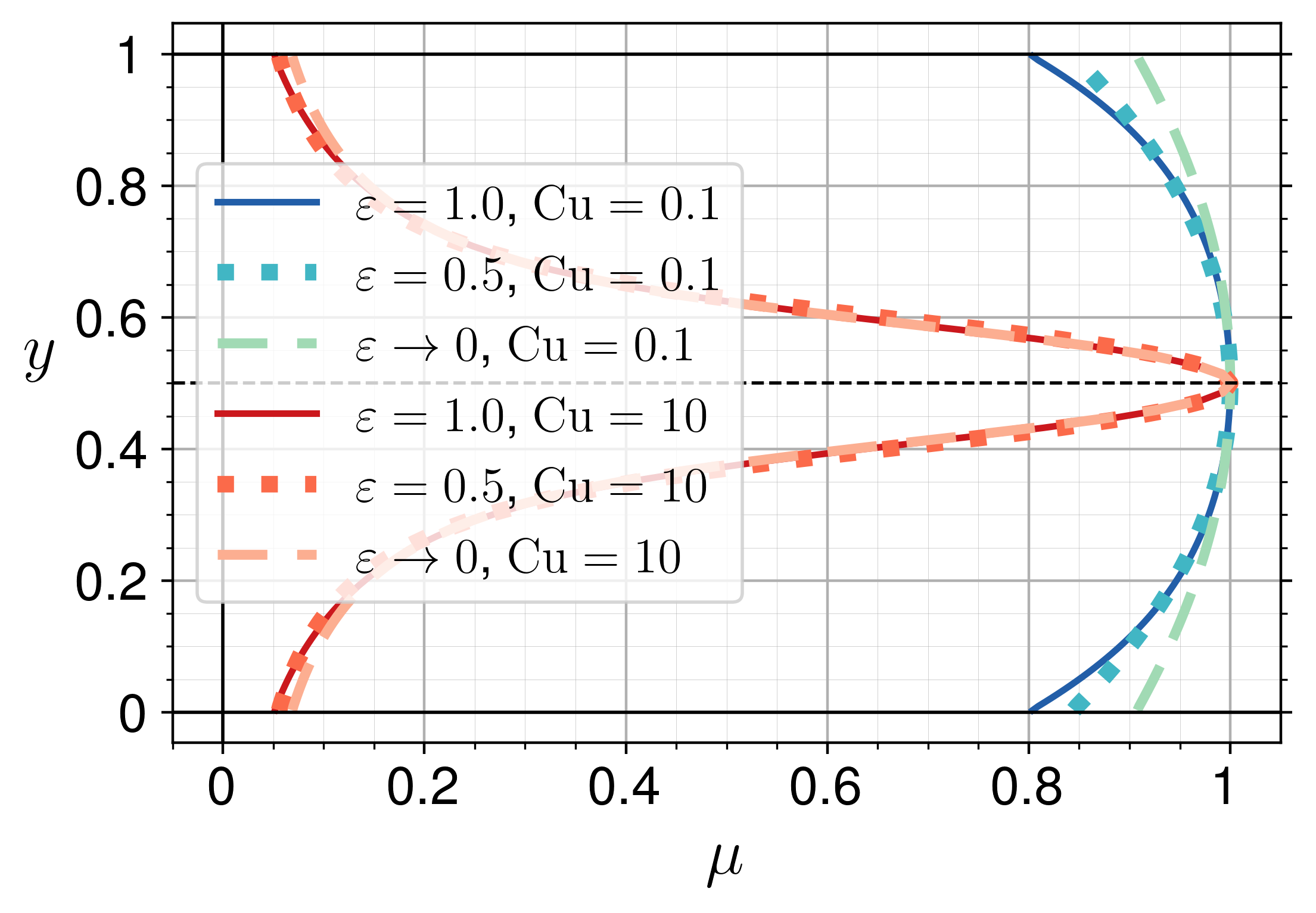}}
	\caption{\label{Fig: Centerline_profiles}Profiles of the velocity (a) and viscosity (b) on the vertical centerline for a rectangular duct of different aspect ratios and for different Carreau numbers. Properties of the non-Newtonian liquid: $m = 0.000135$, $n=0.402$.} 
\end{figure}
\begin{figure}[h!]
	\centering    
	\subfloat[$u$, $n=0.1$]{\includegraphics[width=0.31\textwidth,clip]{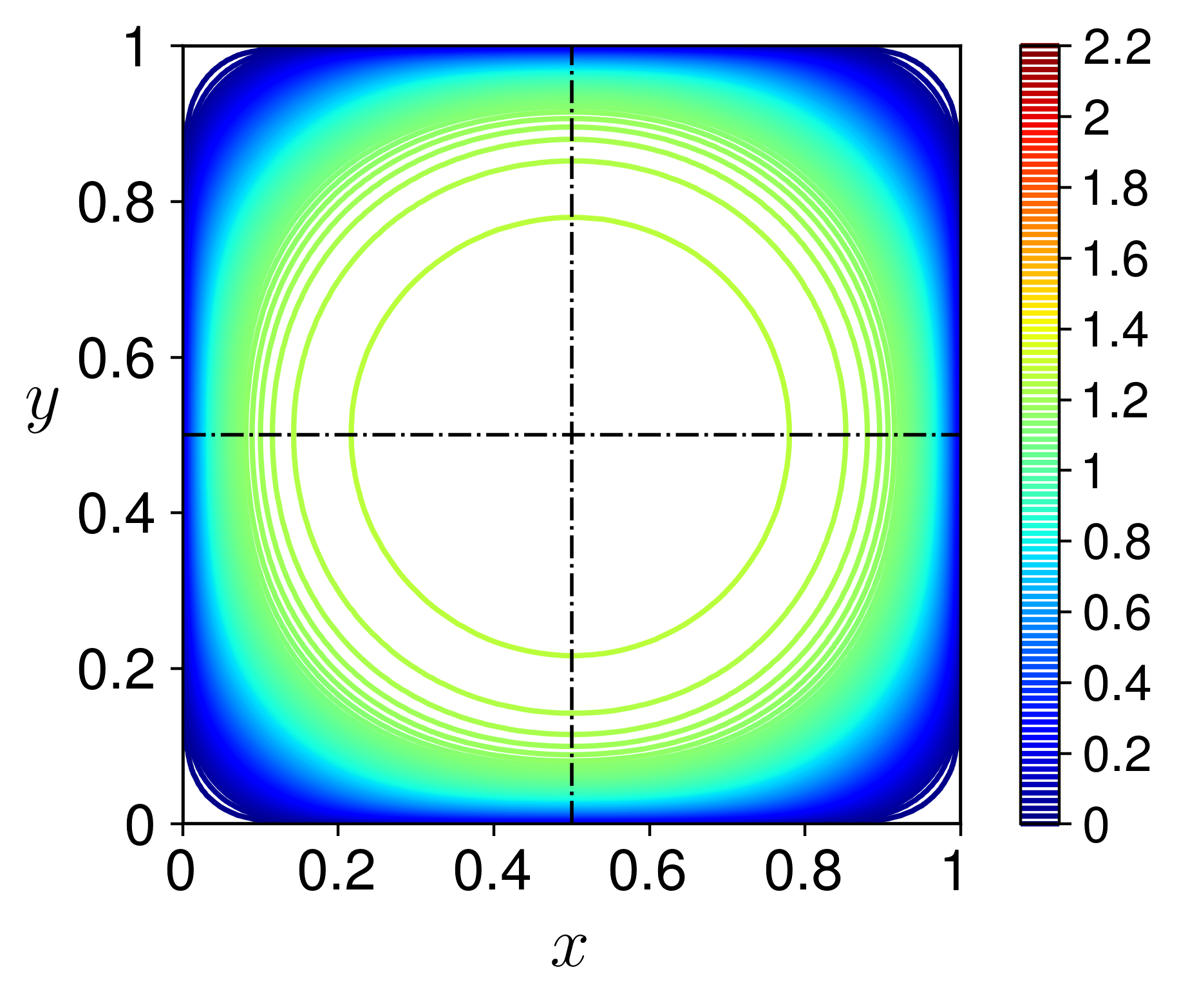}}
	\subfloat[$u$, $n=0.8975$]{\includegraphics[width=0.31\textwidth,clip]{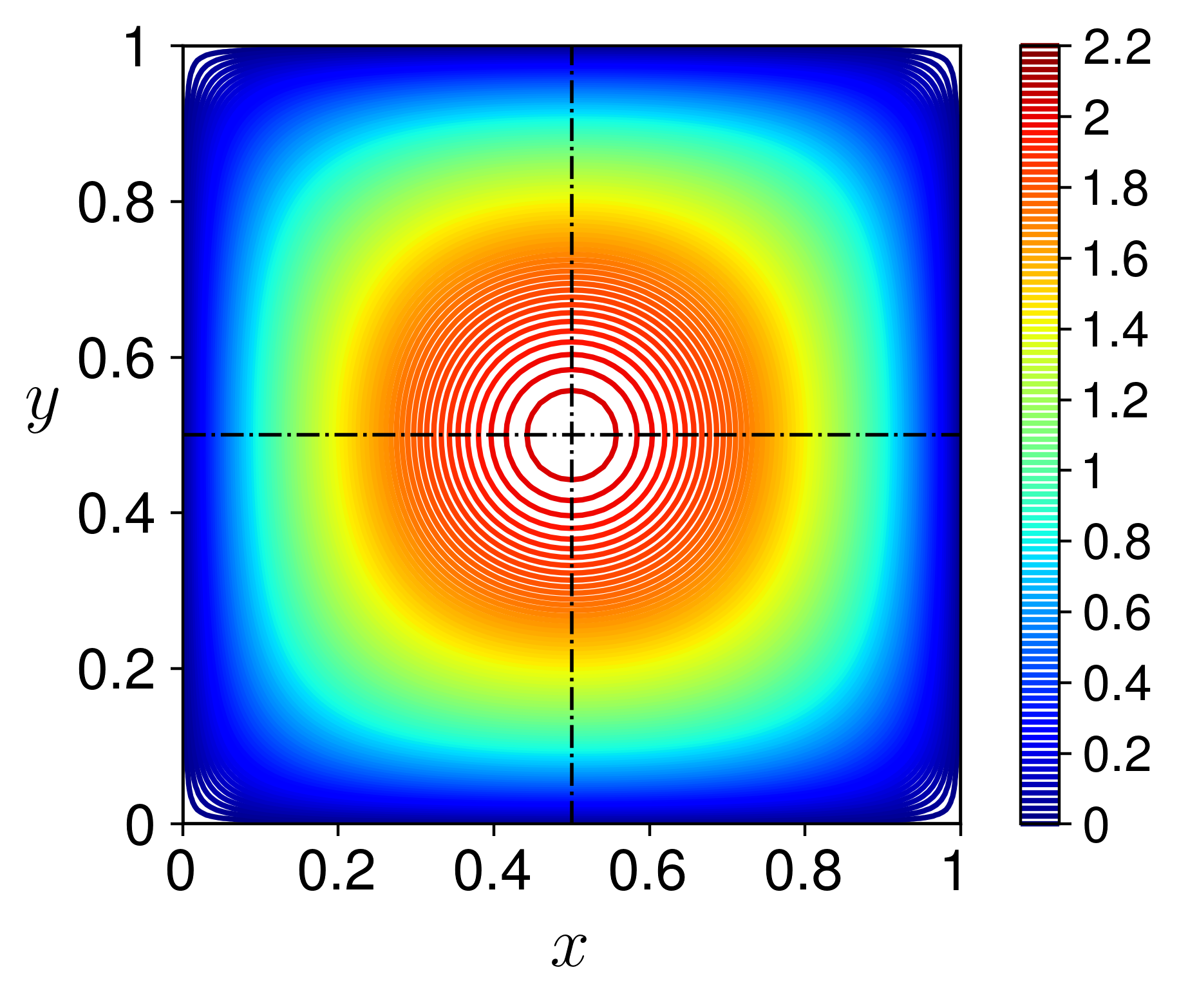}}
	\subfloat[$u$, $x=0.5$]{\includegraphics[width=0.38\textwidth,clip]{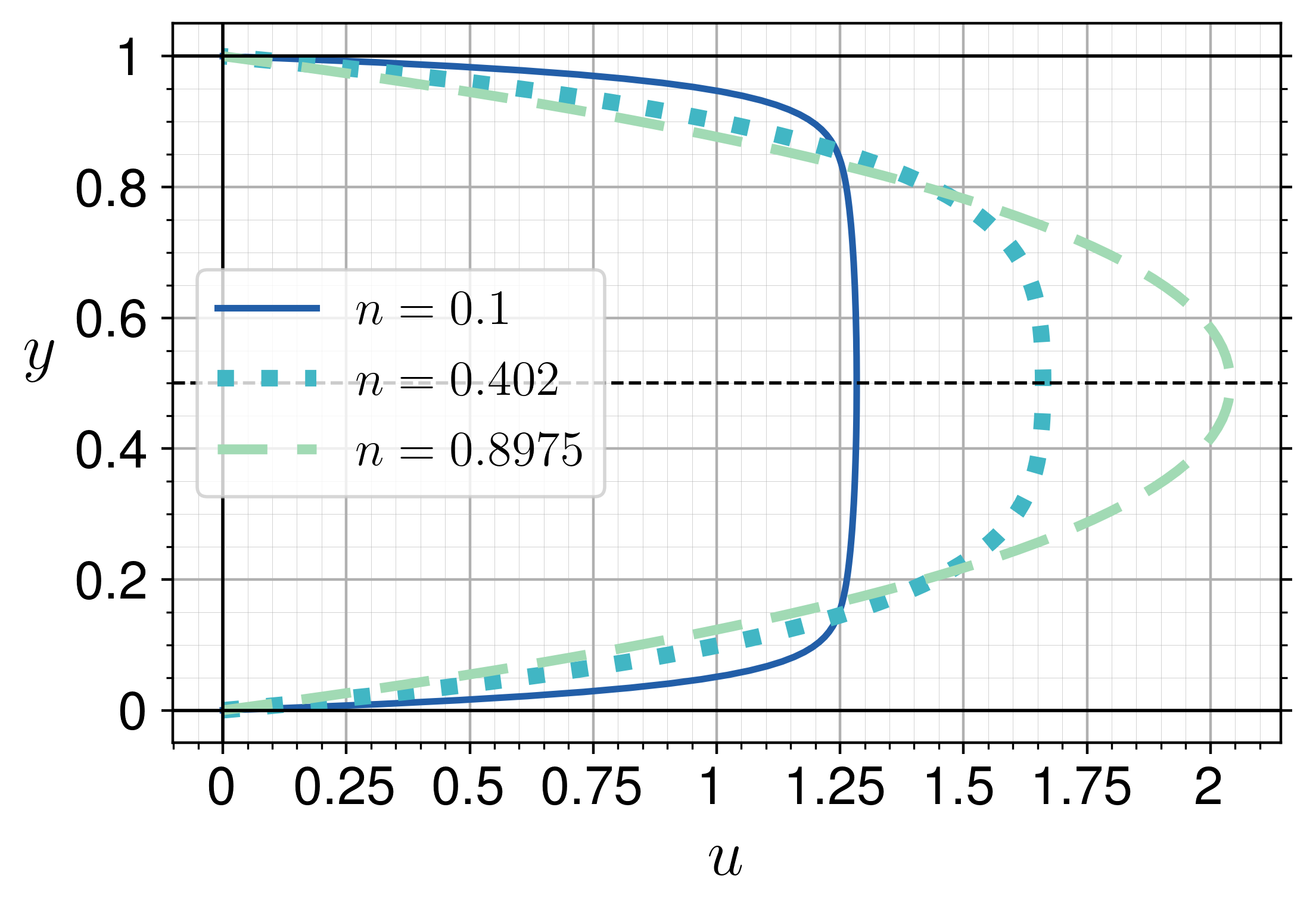}}
	\\
	\subfloat[$\mu$, $n=0.1$]{\includegraphics[width=0.31\textwidth,clip]{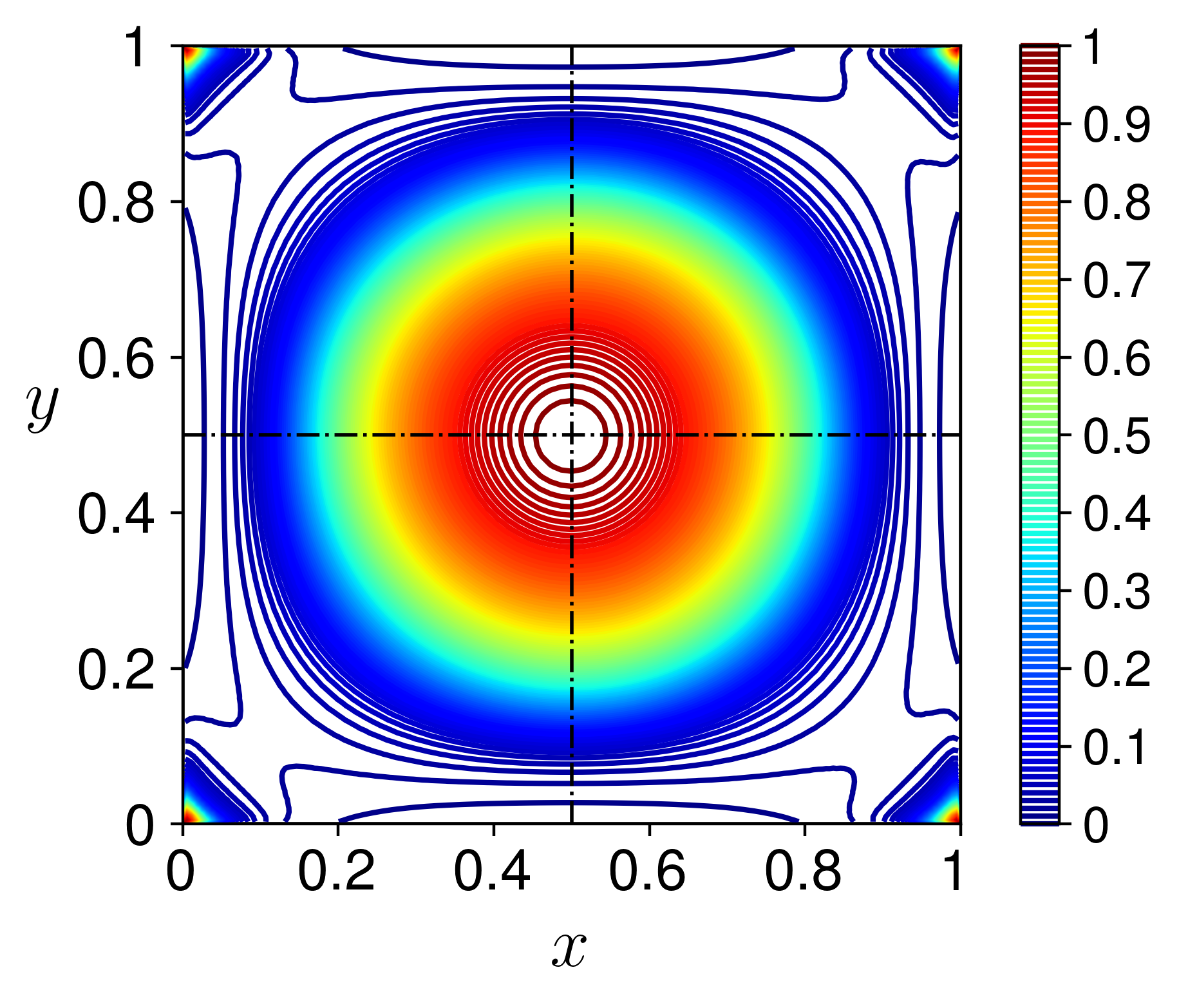}}
	\subfloat[$\mu$, $n=0.8975$]{\includegraphics[width=0.31\textwidth,clip]{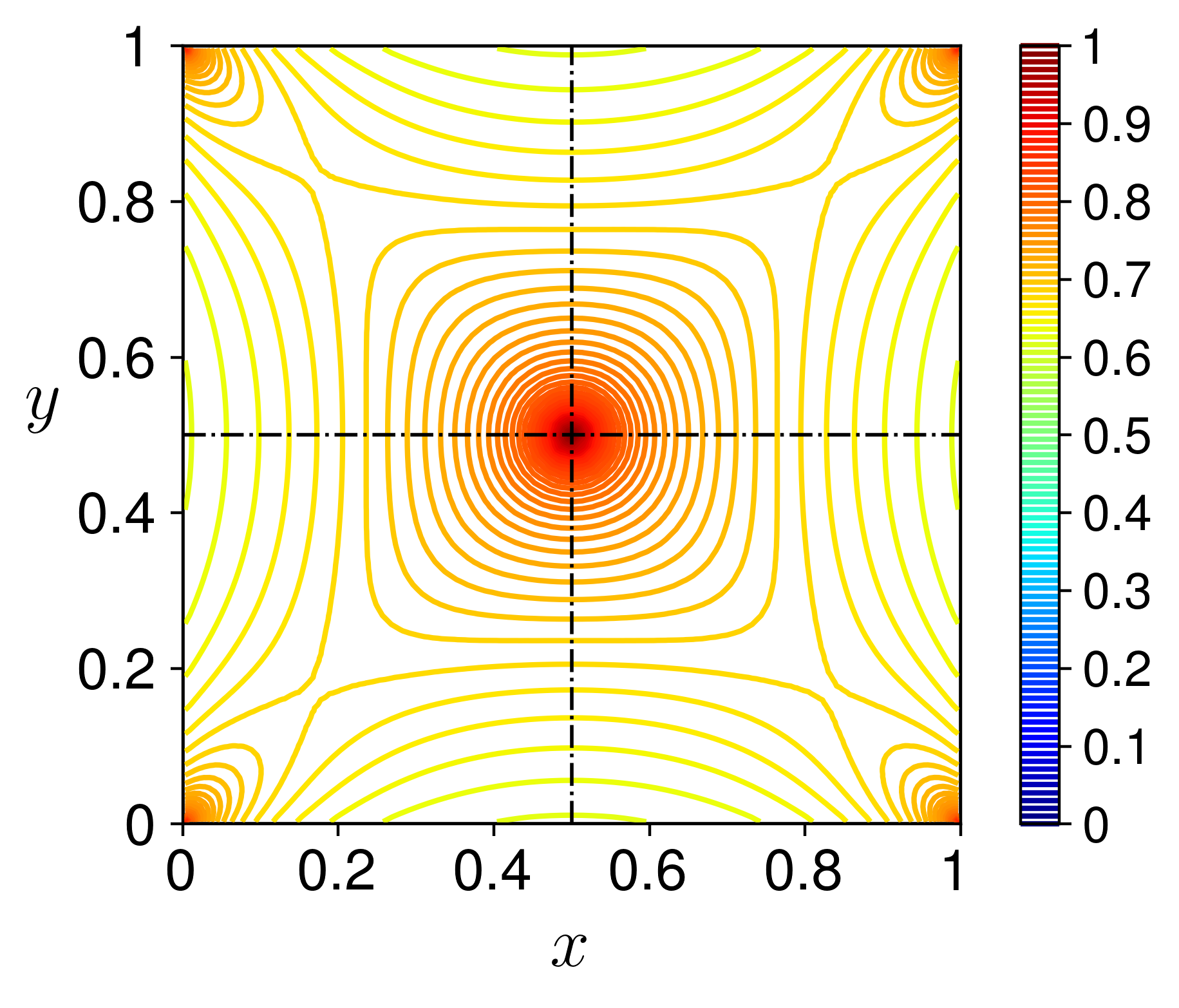}}
	\subfloat[$\mu$, $x=0.5$]{\includegraphics[width=0.38\textwidth,clip]{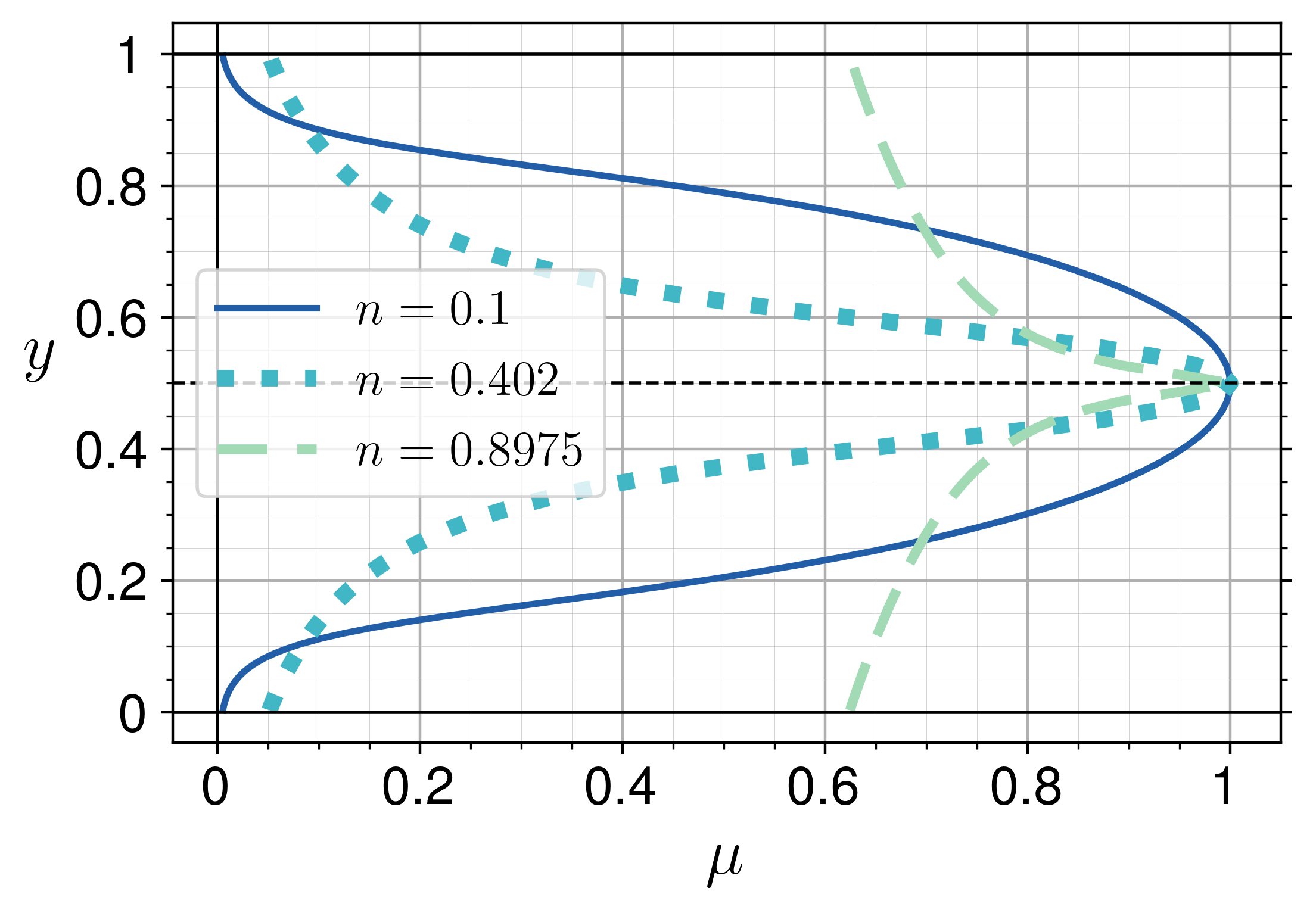}}
	\caption{\label{Fig: Effect_n}Contours and vertical centerline profiles of the velocity (a)-(c) and viscosity (d)-(f) for $m = 0.000135$ and $\Cu=10$ in a square duct. Effect of the shear-thinning index, $n$.} 
\end{figure}

To investigate the effect of the aspect ratio, we show in Fig.\ \ref{Fig: Centerline_profiles}a the velocity and viscosity profiles on the vertical centerline, where the velocity as a function of the $y$-coordinate is depicted for three aspect ratios for small ($\rm \Cu=0.1$) and intermediate ($\rm \Cu=10$) Carreau numbers. The effect of $\rm Cu$ for the same geometry (same line styles in Fig.\ref{Fig: Centerline_profiles}), but different rheological parameters ($n$, $m$), follows the same trends discussed for the two-plate limit presented in Sec.\ \ref{Sec: Results_a}. Here, the purpose is to compare the profiles by varying $\varepsilon$ only, while keeping $n$, $m$, and $\Cu$ fixed. As shown in Fig.\ \ref{Fig: Centerline_profiles}a, the largest velocity is reached at the square duct, and the larger is the duct width (smaller $\varepsilon$, from solid to dotted to dash lines), the smaller is the value of the velocity across the whole centerline, which is a result of the constrain on the average velocity in the flow cross section. The centerline velocity is smaller in wider channels, while the shear rate is less affected by the channel aspect ratio. Consequently, the viscosity is shown to be only slightly dependent on $\varepsilon$, while the effect of $\Cu$ is much more significant (Fig.\ \ref{Fig: Centerline_profiles}b, compare blue and red curves). More specifically, reducing $\varepsilon$ results in minor increase of $\mu$ next to the walls in the power-law region ($\Cu=10$, red curves) and in a slightly larger increase for small $\Cu$ (e.g., $\Cu=0.1$).

Fig.\ \ref{Fig: Effect_n} shows effect of the shear-thinning index, $n$, on the velocity field and the distribution of the non-Newtonian viscosity in the power-law region ($\Cu=10$). A liquid with low $n=0.1$ (Fig.\ \ref{Fig: Effect_n}a,d) exhibits clear shear-thinning behavior across the whole cross section with milder gradients along the diagonal than on the centerlines and flattening of the velocity profile in the center, where $u_{\max}\approx1.28$ compared to $2.1$ in the Newtonian case. This is manifested by the viscosity distribution covering all the values between the two Newtonian limits of $0$ and $1$ in the flow cross section (Fig.\ \ref{Fig: Effect_n}d,f). On the other hand, for a liquid with high $n=0.8975$, which is close to Newtonian, the viscosity varies between $0.6$ and $1$ in the flow cross section (Fig.\ \ref{Fig: Effect_n}e,f). Although the viscosity reaches $\mu=1$ only in the vicinity of the center, the velocity profile is similar to the Newtonian parabolic shape along the centerline with the shear-thinning behavior being important toward the duct walls (Fig.\ \ref{Fig: Effect_n}a,c).
\begin{figure} 
    \centering
        \includegraphics[width=0.6\textwidth,clip]    
        {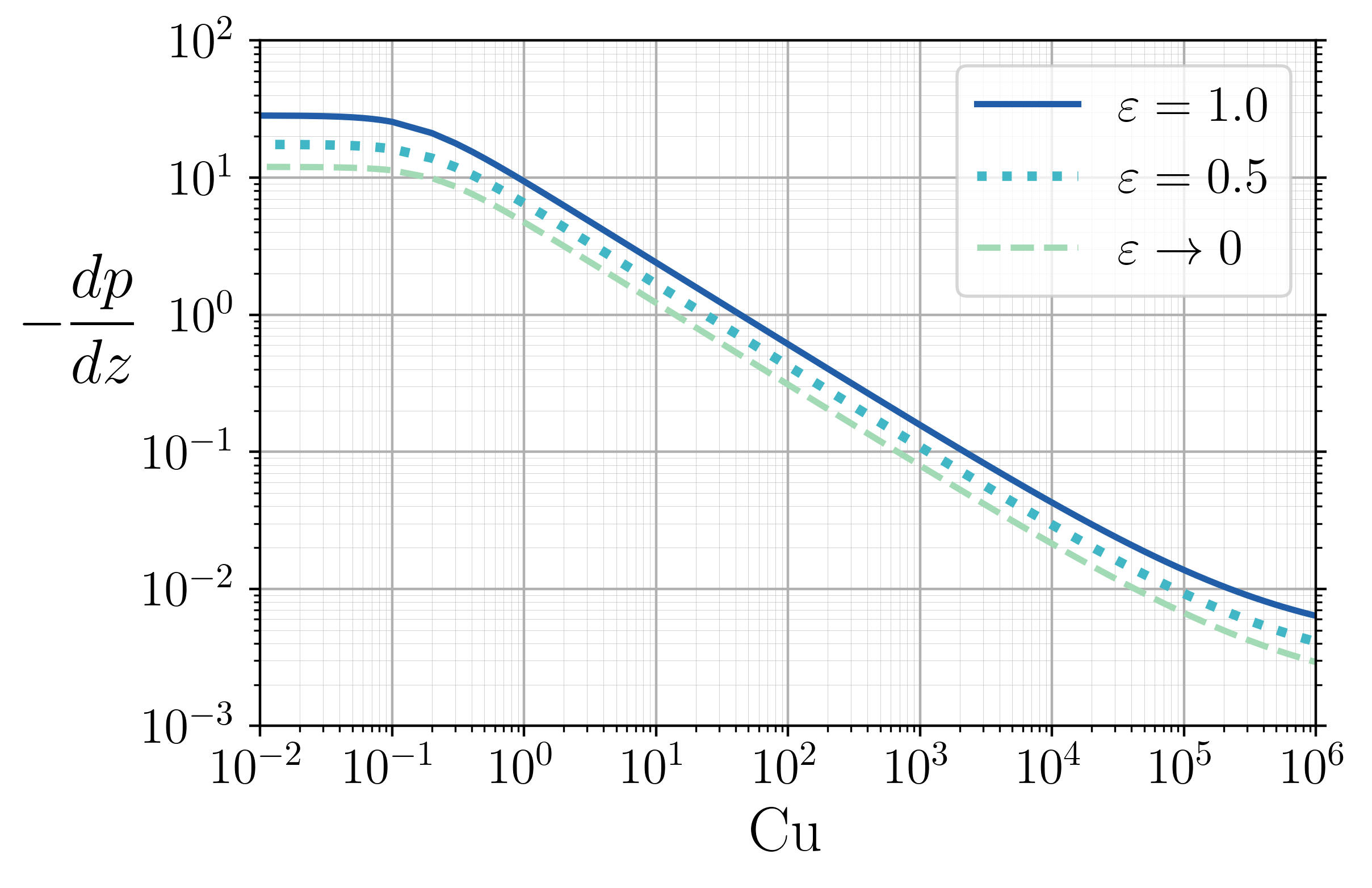}        
    \caption{\label{Fig: Press_grad}Dimensionless pressure gradient as a  function of Carreau number for xanthan gum solution ($m=0.000135$, $n=0.402$) in rectangular ducts of different height-to-width aspect ratios.}
\end{figure}

The aspect ratio is found to have a strong impact on the pressure gradient curve (Fig.\ \ref{Fig:  Press_grad}). For all $\varepsilon$, the shape of the $(-dp/dz)$--$\Cu$ curves are similar to that obtained in the limit of the two-plate geometry (see Sec.\ \ref{Sec: Results_a}). Namely, a clearly observed zero-shear-rate limit, power-law decline of $-dp/dz$ for intermediate $\Cu$, and a large $\Cu$-transition to an infinite-shear-rate limit is observed for $\Cu\gg1$. The curves in Fig.\ \ref{Fig:  Press_grad} shift up and to the right with increasing $\varepsilon$, suggesting that the flow through a duct of a smaller width requires a higher pressure gradient. In the limit of a Newtonian liquid, i.e., for $\Cu= 0$, the well-established analytical results of \citet{Cornish28}, $-dp/dz\approx17.48$ and $\approx28.43$, are retrieved for $\varepsilon=0.5$ and $1$, respectively. Increasing $\Cu$ to $0.1$ (i.e., the case that is characterized in Fig.\ \ref{Fig: Contours} as an almost Newtonian flow), results in $-dp/dz$ lower by almost $10\%$, i.e., in a value of $\approx25.53$ in a square duct. Further increase of $\Cu$ to $10$, leads to a pressure gradient of about a tenth of the Newtonian value at $(-dp/dz)\approx2.4$. The same trend is observed for all other aspect ratios. 

The similarity in the $(-dp/dz)$--$\Cu$ relations for various $\varepsilon$ for fixed values of $n$ and $m$ of the Carreau fluid points out to a possibility of generalizations described in the next section.

%%%%%%%%%%%%%%%%%%%%%%%%%%%%%%%%%%%%%%%%%%%%%%%%%%%%%%%%%%%%%%%%%%%%%%%%%%%
\subsection{\label{Sec: Results_c}Generalized scaling for the pressure gradient and friction factor}

For steady fully developed laminar flow in a rectangular duct, the resistance to flow as a result of friction on the walls is usually expressed in terms of a dimensionless friction factor, defined as 
\begin{equation}\label{Eq: f}
    f=\frac{2\bar{\tau}_w}{\rho U^2},
\end{equation}
where $\bar{\tau}_w$ is the dimensional average shear stress acting on the channel walls (see \citet{White11}) given by 
\begin{equation}\label{Eq: tau}
    \bar{\tau}_w=\frac{HL}{2(H+L)}\biggl(-\frac{d\tilde{p}}{d\tilde{z}}\biggr)\, .
\end{equation}
The dimensional pressure gradient, $-d\tilde{p}/d\tilde{z}$, is obtained from the dimensionless pressure gradient, $-dp/dz$, which is computed numerically as discussed in Section \ref{Sec: Numerics}:
\begin{equation}\label{Eq; dp}
     \biggl(-\frac{d\tilde{p}}{d\tilde{z}}\biggr)=\frac{\mu_0U}{H^2}\biggl(-\frac{dp}{dz}\biggr)\, . %({\rm Cu},n,\varepsilon).
\end{equation}
Combining Eqs. (\ref{Eq: f}) with (\ref{Eq: tau}) and (\ref{Eq; dp}), we get the following relation for the friction factor
\begin{equation} \label{Eq: f_2}
    f=\dfrac{(-dp/dz)}{\dfrac{\rho U H}{\mu_0}(1+\varepsilon)}.
\end{equation}
By multiplying and dividing Eq.\ (\ref{Eq: f_2}) by the analytical expression of \citet{Cornish28} for the  dimensionless pressure gradient of Newtonian flow in a rectangular duct  
\begin{equation} \label{Eq: dPdz_N}
    \biggl(-\frac{dp}{dz}\biggr)_{N} = 12 
		 \Bigg[1 - \frac{192\, \varepsilon}{\pi^5}
    \bigg(\tanh\biggl(\frac{\pi}{2\, \varepsilon}\biggr) + \frac{1}{3^5}\tanh\biggl(\frac{3\pi}{2\, \varepsilon}\biggr) + ...\bigg)\Bigg]^{-1}\, ,
\end{equation}
we obtain the following equivalent expression of the friction factor
\begin{equation}\label{Eq: Friction_factor}
    f=\dfrac{12\,\mu_0\,\mathcal{P}}{{\rho\, U\, H (1+\varepsilon)\Bigg[1 - \dfrac{192\, \varepsilon}{\pi^5}
    \bigg(\tanh\biggl(\dfrac{\pi}{2\, \varepsilon}\biggr) + \dfrac{1}{3^5}\tanh\biggl(\dfrac{3\pi}{2\, \varepsilon}\biggr) + ...\bigg)\Bigg]}}\, ,
\end{equation}
where $\mathcal{P}$ is the ratio between the dimensionless non-Newtonian pressure gradient to the Newtonian one with the same average velocity $U$. Specifically, $\mathcal{P}$ is referred to as scaled pressure gradient and is denoted as
\begin{equation} \label{Eq: P}
    \mathcal{P}({\rm Cu},n,m,\varepsilon)
    = {\biggl(-\dfrac{dp}{dz}\biggr)}\bigg/{\biggl(-\dfrac{dp}{dz}\biggr)_{N}}\, .
\end{equation}

\begin{figure}[] 
	\centering
        \includegraphics[width=0.5\textwidth,clip]{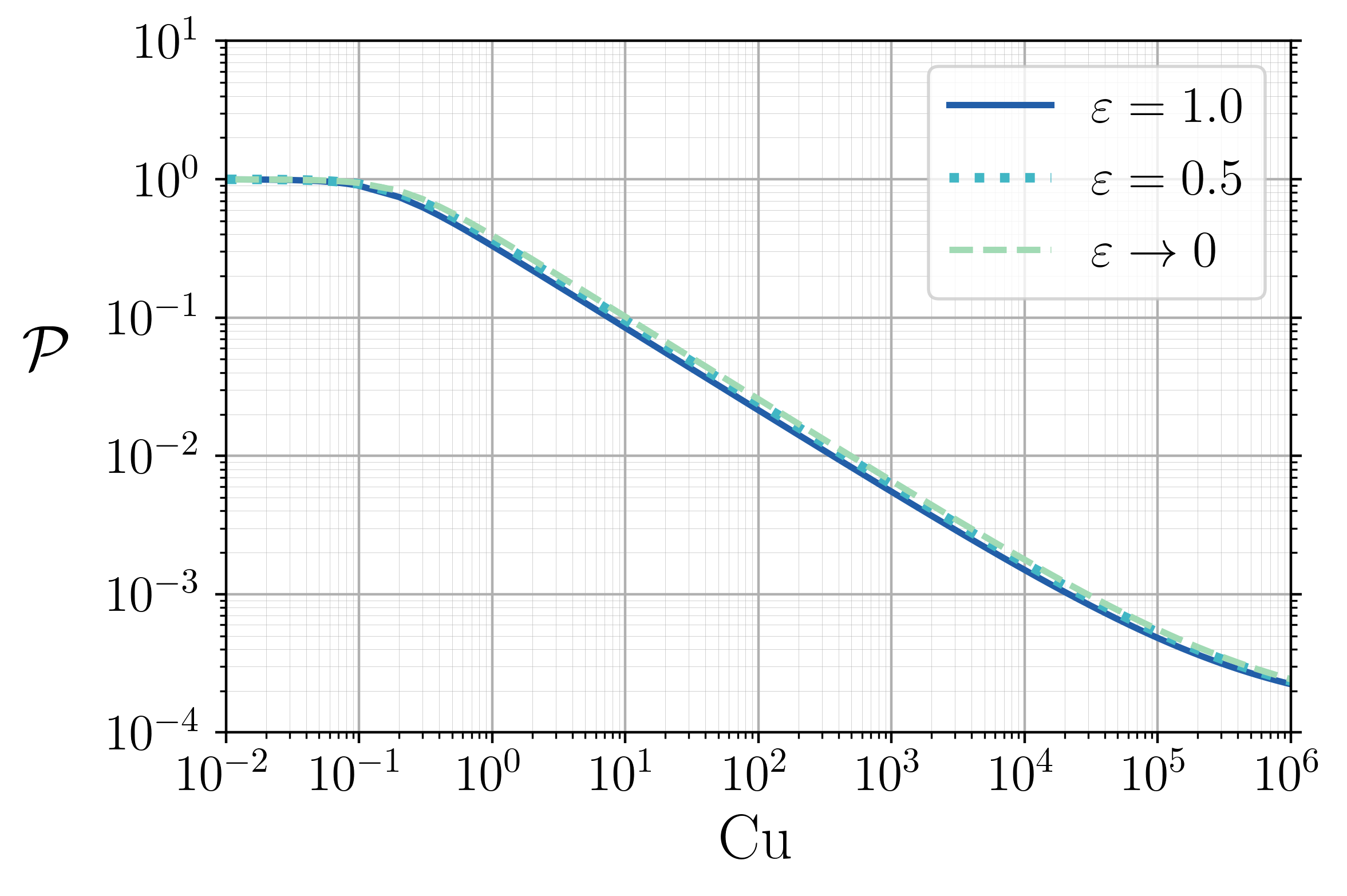}
	\caption{\label{Fig: P_var_AspRatio} Scaled pressure gradient, $\mathcal{P}$ (Eq.\ \ref{Eq: P}), as a function of Carreau number for xanthan gum solution ($m=0.000135$, $n=0.402$) in rectangular ducts of different aspect ratios.}
\end{figure}

Fig.\ \ref{Fig: P_var_AspRatio} shows the scaled pressure gradient for the flow of xanthan gum solution in rectangular ducts (data shown in Fig.\ \ref{Fig: Press_grad}). One can observe that the difference in the curves obtained for various aspect ratios shrinks to a rather slight horizontal shift for intermediate $\Cu$ in the power-law region (i.e., up to $15\%$ difference in $\Cu$ for the same $\mathcal{P}$, note the logarithmic coordinates), while the same Newtonian limit of $1$ is reached for $\Cu\ll1$ (as expected from the definition of $\mathcal{P}$). This suggests that the effects of the channel geometry and rheological parameters may be decoupled: the scaled pressure gradient captures well the effect of the shear-thinning rheology on the pressure gradient but another rescaling is needed to account for the effect of the aspect ratio. By inspection of Eq. (\ref{Eq: Friction_factor}), we can define the effective channel size $H_e$ as the rectangular duct equivalent to the two-plate height $H$:
\begin{equation}
      H_{e} = H \bigg(1 + \varepsilon\bigg) 
    \Bigg[1 - \frac{192\, \varepsilon}{\pi^5}
    \bigg(\tanh\biggl(\frac{\pi}{2\, \varepsilon}\biggr) + \frac{1}{3^5}\tanh\biggl(\frac{3\pi}{2\, \varepsilon}\biggr) + ...\bigg)\Bigg].
\end{equation}
According to this definition, $H_e(\varepsilon)$ is a function of $\varepsilon$  and it differs from the channel height for all rectangular ducts, except for the two-plate limit, $\varepsilon \to 0$, where $H_e = H$. Then, we can also define an effective viscosity, $\mu_e$, as the deviation from the Newtonian viscosity $\mu_0$ 
\begin{equation}
    \mu_e(\Cu,n,m,\varepsilon)=\mu_0 \, \mathcal{P},
\end{equation}
and rewrite Eq.\ (\ref{Eq: Friction_factor}) in terms of an effective Reynolds number based on the effective size and viscosity
\begin{equation}
    \Rey_e=\frac{\rho H_e U}{\mu_e},
\end{equation}
With these definitions, the friction factor can be reformulated as a function of the (effective) Reynolds number only, whereby similarly to Newtonian fluids 
\begin{equation}
    f=\frac{12}{\Rey_e}.
\end{equation}
Since the effective viscosity depends on the Carreau number (Eq.\ \ref{Eq: Cu}), an effective Carreau number, which is based on the effective channel size, is introduced:
\begin{equation}
    \Cu_{e} = \frac{\lambda U}{H_{e}}.
\end{equation}

\begin{figure}[] 
	\centering
        \subfloat[$n = 0.1$]{\includegraphics[width=0.48\textwidth,clip]{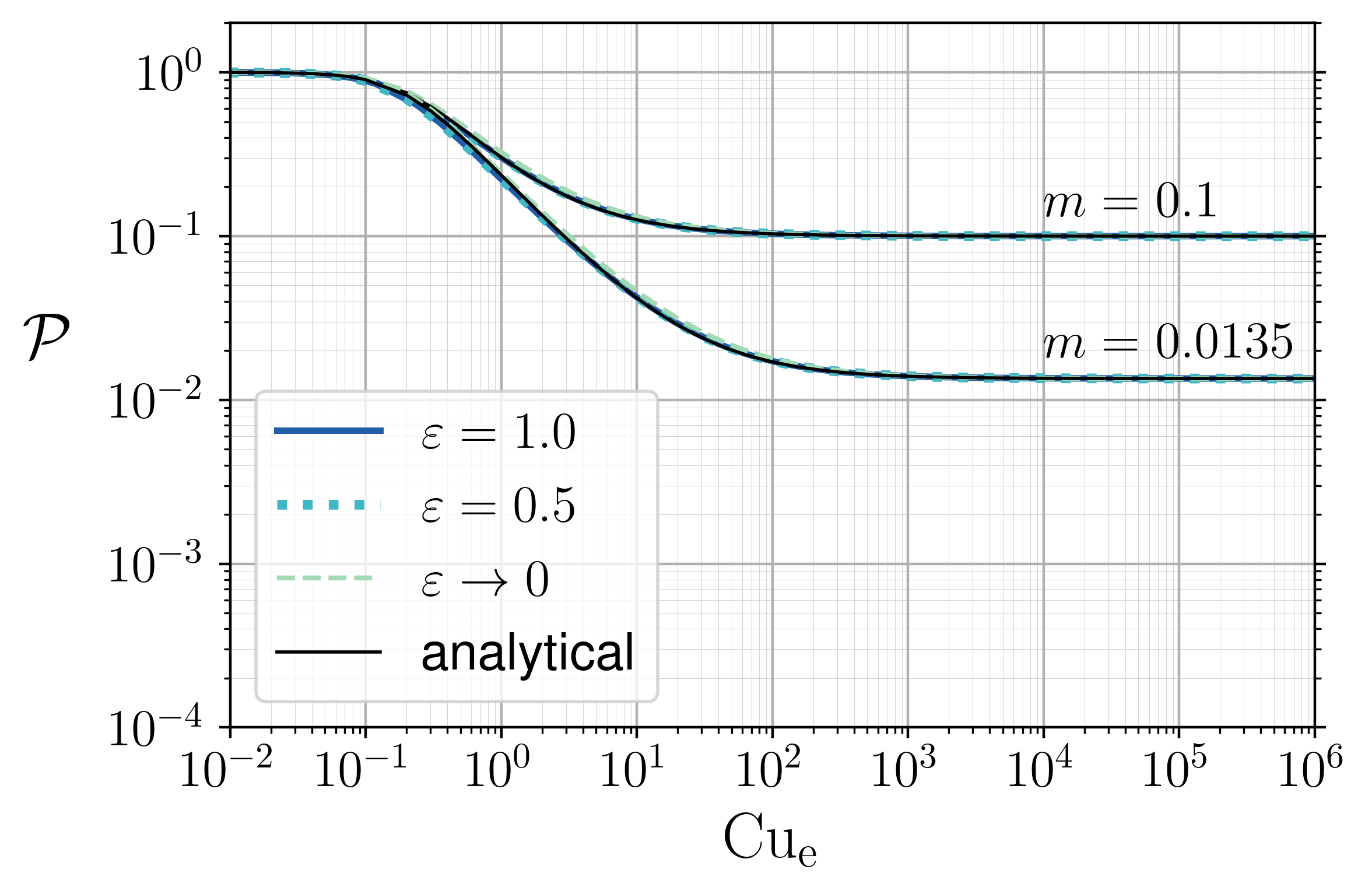}}
        \subfloat[$m = 0.000135$]{\includegraphics[width=0.48\textwidth,clip]{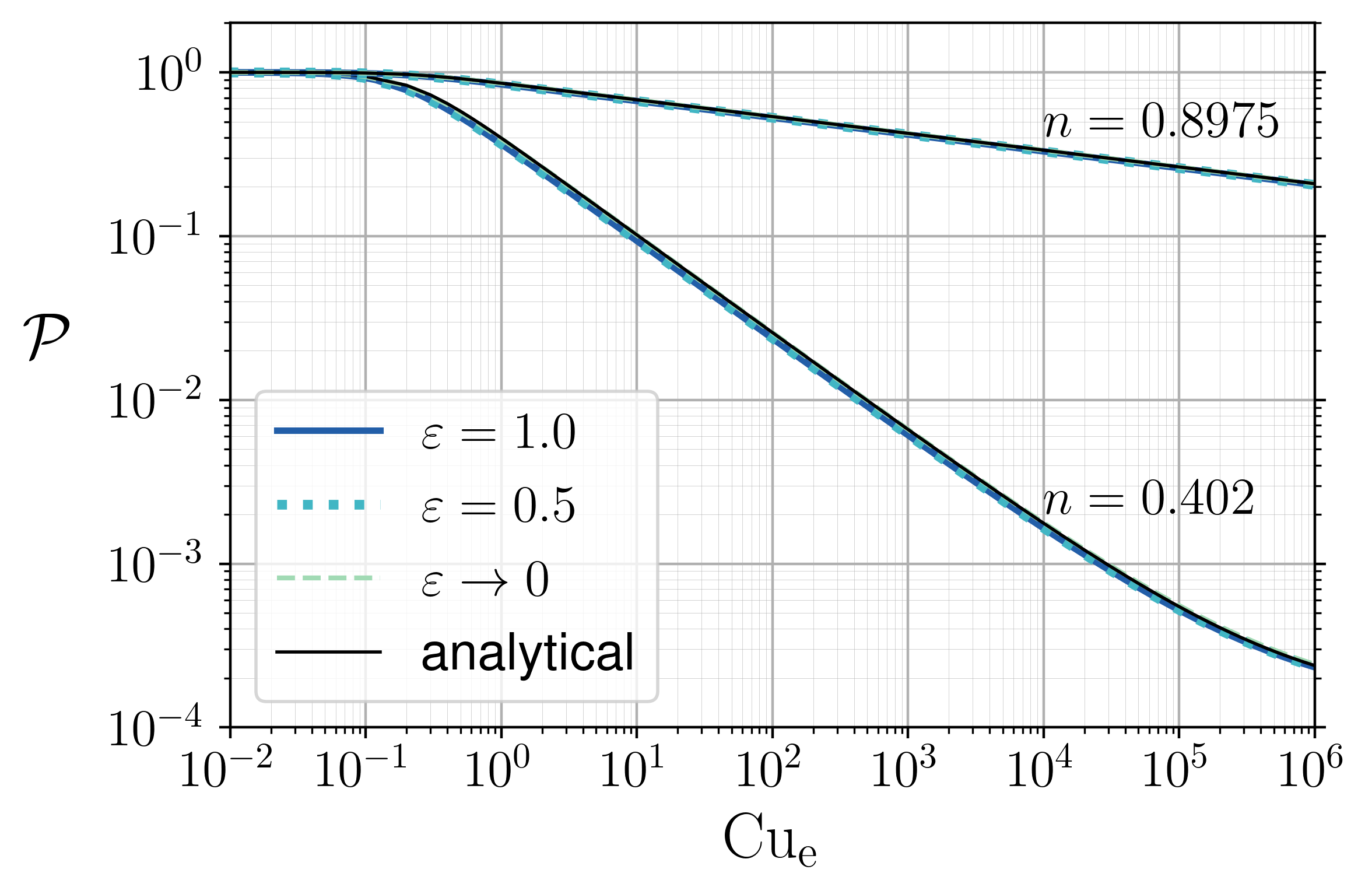}}
        \caption{\label{Fig: P_numerical_vs_analytical} Scaled pressure gradient, $\mathcal{P}$, as a function of the effective Carreau number. Comparison of numerical results for rectangular ducts of different aspect ratios with analytically obtained fitting function (solid black line, Eq.\ (\ref{Eq: P_Acu})).}
\end{figure}
\begin{figure}[]
    \centering    
    \subfloat[Pressure coefficient]{\includegraphics[width=0.48\textwidth,clip]{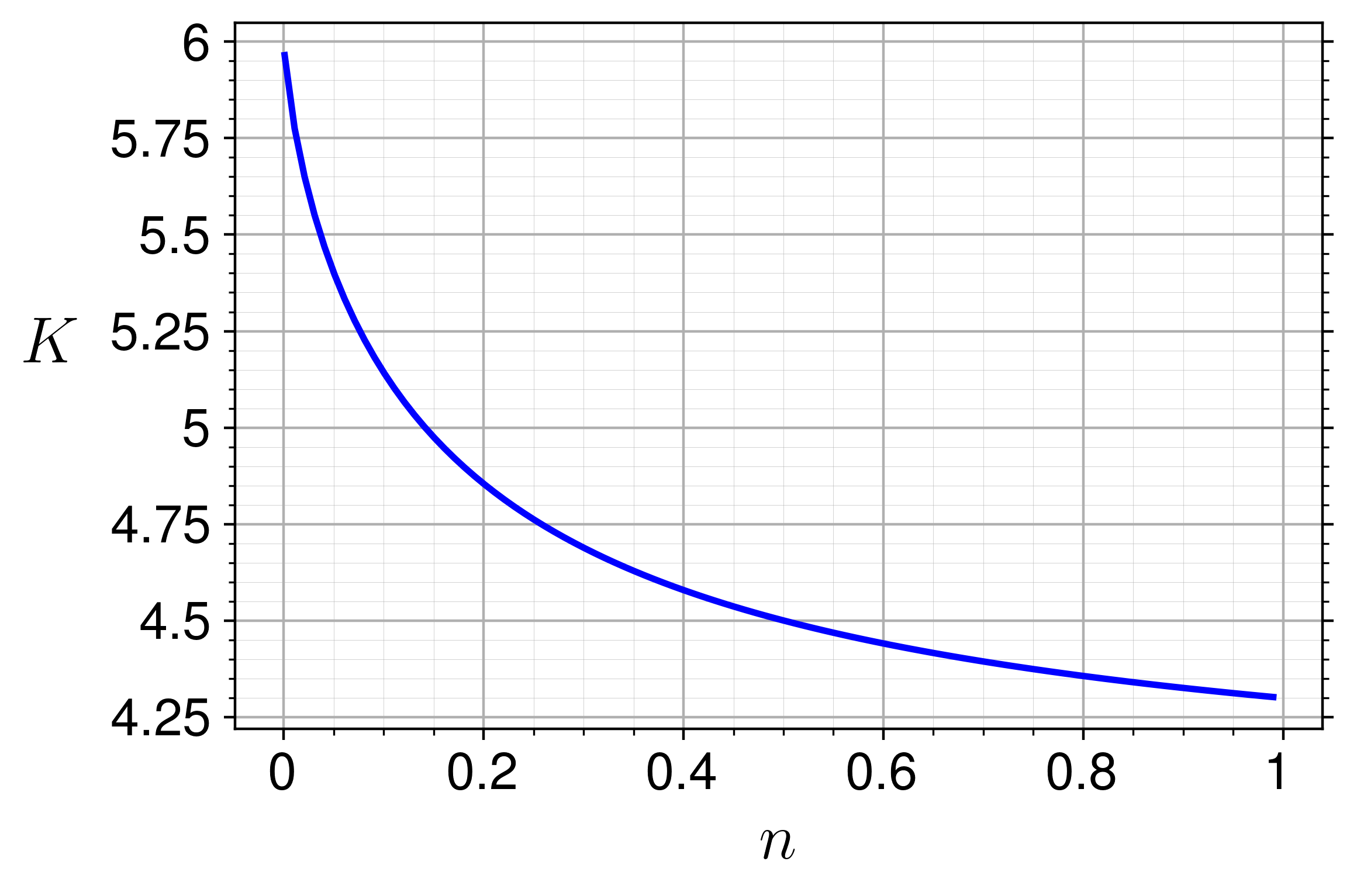}}
    \subfloat[Scaled pressure gradient]{\includegraphics[width=0.48\textwidth,clip]{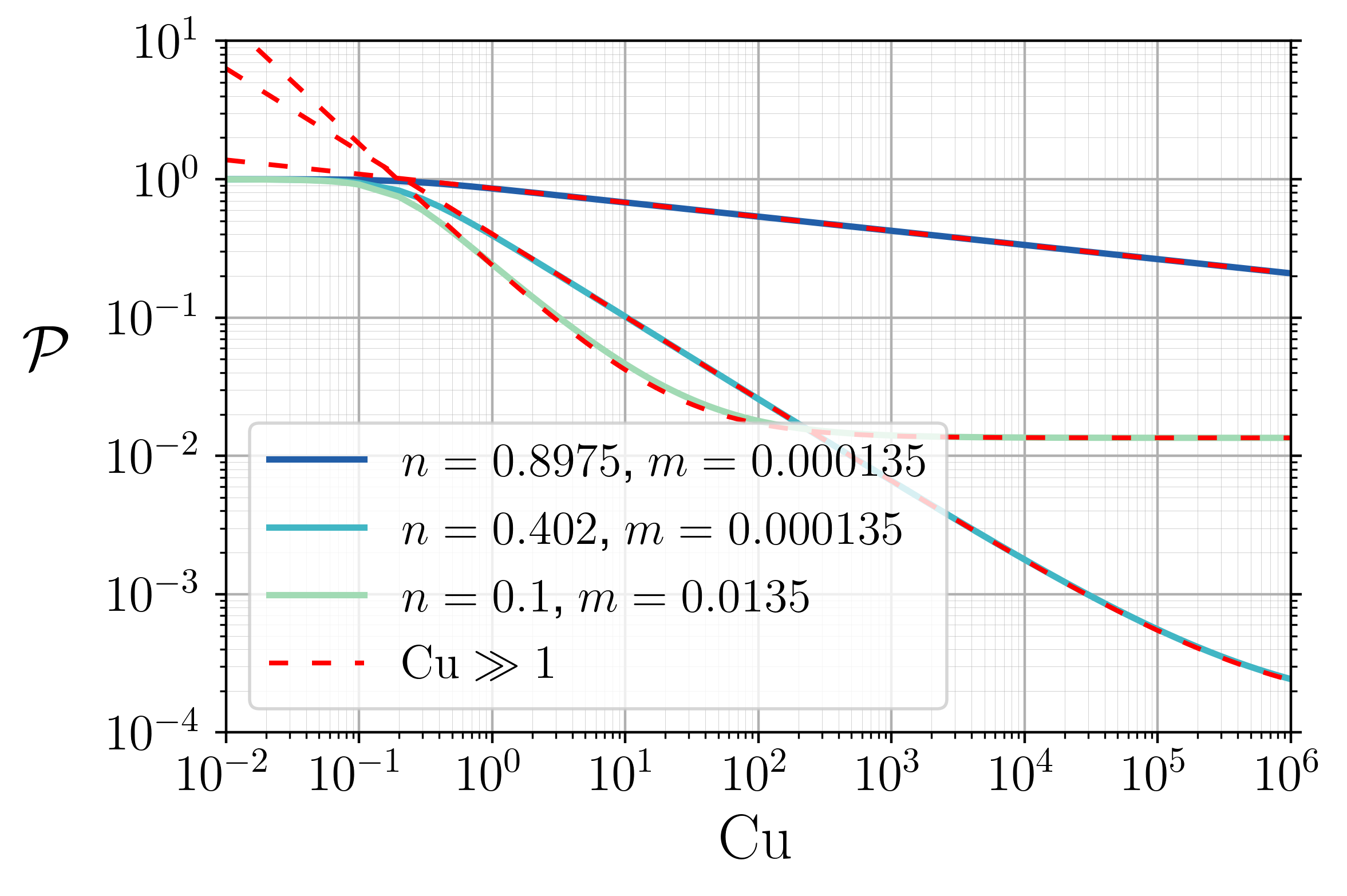}}    
	\caption{\label{Fig: Large_Cu_limit}(a) Dependence of coefficient $K$ on the shear-thinning index $n$. (b) Comparison of numerical (solid lines) and analytical (dashed lines, the limit of $\Cu\gg1$) dependencies of $\mathcal{P}$ on $\Cu$ for the two-plate geometry, $\varepsilon\to0$.}
\end{figure}

As shown in Fig.\ \ref{Fig: P_numerical_vs_analytical}, once $\mathcal{P}$ is recomputed as a function of $\Cu_e$, the $\mathcal{P}$--$\Cu$ curves for different $\varepsilon$ shown in Fig.\ \ref{Fig: P_var_AspRatio} collapse to the same curve, indicating that $\mathcal{P}$ becomes independent on the aspect ratio and depends only on the rheological parameters ($m$ and $n$) and $\Cu_{e}$. Moreover, the $\mathcal{P}$--$\Cu_e$ curve can be represented by an explicit equation, which is similar to the Carreau fluid rheology equation (cf. Eq. \ref{Eq: Mu})
\begin{equation} \label{Eq: P_Acu}
    \mathcal{P}
    = m 
    + \bigg(1 - m\bigg)
    \Biggl(1
    + \biggl[
    K \Cu_{e}
    \biggr]^2 \Biggr)^{(n-1)/2},
\end{equation}
where the coefficient $K(n)$ depends only on the shear-thinning index $n$. The value of $K$ can be found from the analytical solution in the two-plate limit for $\displaystyle\Cu\gg1$ (see detailed derivation in Appendix \ref{Sec: Large_Cu}). The following relation for the rescaled pressure drop, considering, without loss of generality, a fluid with $m=0$, is obtained:
\begin{equation} \label{Eq; P_Cu_gg_1}
    \mathcal{P}(\Cu\gg1,n,m=0,\varepsilon\to0)
    = \frac{2^{n-1}}{3}
    \biggl[\frac{2n+1}{n}\biggr]^n
    \Cu^{n-1}.
\end{equation}
Comparing Eq.\ (\ref{Eq; P_Cu_gg_1}) with the large-Carreau-limit of Eq.\ (\ref{Eq: P_Acu})
\begin{equation}
    \mathcal{P}(\Cu\gg1,n,m=0,\varepsilon\to0)
    = K^{n-1}
    \Cu^{n-1},
\end{equation}
yields the following analytical expression for $K$
\begin{equation} \label{Eq: A_Cu}
    K 
    = \frac{2}{3^\frac{1}{n-1}}\biggl[\frac{2n+1}{n}\biggr]^\frac{n}{n-1}.
\end{equation}
The variation of $K$ with $n$ is presented in Fig.\ \ref{Fig: Large_Cu_limit}a. Since for shear-thinning liquids $n$ varies between $0$ and $1$, two limiting values of $k$ can be identified as
\begin{equation*}
    \lim_{n\to0} \biggl(\frac{2}{3^\frac{1}{n-1}}\biggl[\frac{2n+1}{n}\biggr]^\frac{n}{n-1}\biggr) = 6,
    \qquad
    \qquad
    \lim_{n\to1} \biggl(\frac{2}{3^\frac{1}{n-1}}\biggl[\frac{2n+1}{n}\biggr]^\frac{n}{n-1}\biggr) = \frac{6}{e^{1/3}}
    \approx 4.299.
\end{equation*}
Although Eq.\ (\ref{Eq: A_Cu}) is obtained for $m=0$, the same expression for $K$ is recovered for any arbitrary value of $m$. This can be validated by comparison between the $\Cu\gg1$ limit for $\mathcal{P}$ (with $K$ from Eq.\ \ref{Eq: A_Cu}) and the exact numerical results obtained for the same parameters (Fig.\ \ref{Fig: Large_Cu_limit}b). For $m>0$, a good agreement, i.e., overlapping curves, is found not only for very large values of $\Cu$, but also for the whole region of nearly Newtonian $\mu=m$ viscosity (see the curves for $n=0.1$ for $\Cu>10^3$) and for the power-law region of moderate $\Cu$. Obviously, the analytical asymptote for $\Cu\gg1$ does not predict the proper behavior for small $\Cu$ ($<0.1$). Nevertheless, once the analytically obtained $K$ (Eq.\ \ref{Eq: A_Cu}) is introduced into the expression for $\mathcal{P}$ (Eq.\ \ref{Eq: P_Acu}), the latter fits well with the numerical results also for small $\Cu$.

To sum up, the introduction of the rescaled pressure gradient $\mathcal{P}$ and the effective Carreau number, $\Cu_{e}$, results in a collapse of the $\mathcal{P}$--$\Cu_{e}$ curves for all aspect ratios, see Fig.\ \ref{Fig: P_numerical_vs_analytical}. In fact, the proposed scaling for the pressure gradient curve seems to be universal for all fluids whose rheology is defined by the Carreau model (Eq.\ \ref{Eq: Mu}) and flowing in rectangular ducts of an arbitrary aspect ratio. Accordingly, the effective Reynolds number (and the resulting friction factor) can be calculated based on the effective viscosity, given by
\begin{equation}
    \mu_{e}
    = \mu_0 
    + \bigg(\mu_0 - \mu_\infty\bigg)
    \Biggl(1
    + \biggl[
    K(n) \Cu_{e}
    \biggr]^2 \Biggr)^{(n-1)/2}.
\end{equation}

%%%%%%%%%%%%%%%%%%%%%%%%%%%%%%%%%%%%%%%%%%%%%%%%%%%%%%%%%%%%%
\section{Conclusions}\label{Sec: Conclusions} 

A rigorous solution for a steady laminar flow in a rectangular duct is found numerically for a shear-thinning fluid, whose viscosity is described by the Carreau model. The two-dimensional profiles of the velocity field and viscosity are presented for a wide range of Carreau numbers, including in the power-law region. The effects of both the fluid rheology and the duct aspect ratio on the velocity field and integral characteristics of the flow, such as the pressure gradient, are addressed and discussed in detail.

Our analysis shows that both the rheology and channel geometry noticeably affect the pressure gradient dependence on the Carreau number, which is the dimensionless parameter of the non-Newtonian viscosity model. Separate consideration of these contributions allows us to derive universal scaling laws for the effective viscosity and the effective channel size, which generalize the classical formula for the friction factor of Newtonian flows to shear-thinning fluids flowing in a rectangular channel of an arbitrary aspect ratio. Specifically, the friction factor can still be computed using the classical $12/\rm Re_e$ relation, where the Reynolds number is based on an effective length scale that depends on the aspect ratio only and the effective viscosity is a function of the rheology dimensionless parameters. These findings are supported by comparison with numerical results. Our results clarify the effect of the fluid rheology and channel aspect ratio on frictional pressure gradients and may inspire the modeling of flow of shear-thinning fluids in complex  geometries.

\begin{acknowledgments}
    This research was supported by Israel Science Foundation (ISF) grant No 415/18.
\end{acknowledgments}

%%%%%%%%%%%%%%%%%%%%%%%%%%%%%%%%%%%%%%%%%%%%%%%%%%%%%%%%%%%%%%%%%%%%%%%%%%%
\appendix

\section{Rescaled pressure gradient in the large-Carreau-number limit} \label{Sec: Large_Cu}

In the two-plate limit ($u=u(y)$, $y\in[0,1]$), the shear stress in the limit of $\Cu\gg1$ can be written in case of $m=0$ as
\begin{equation}
    \tau_{y z} 
    = \biggl(1 
    + \Cu^2 \biggl[
    \frac{d u}{d y} 
    \biggr]^2\biggr)^{(n-1)/2}
    \frac{d u}{d y}
    = \Cu^{n-1} 
    \biggl(\frac{d u}{d y}\biggr)^n
\end{equation}		
and the momentum equation, Eq.\ (\ref{Eq: Momentum}), reduces to
\begin{subequations}
    \begin{align}
        n \Cu^{n-1}
        \biggl(\frac{d u}{d y}\biggr)^{n-1}
        \frac{d^2 u}{d y^2}
        &= \frac{d p}{d z}
    \end{align}
\end{subequations}
Imposing the no-slip boundary condition on the wall at $y=0$ and the symmetry condition in the middle at $y=0.5$, the momentum equation can be integrated to get an expression for the velocity profile 
\begin{equation}
        u(y,\Cu,n) 
        = 
        \Cu^\frac{1-n}{n}
            \biggl(-\frac{d p}{d z}\biggr)
            ^\frac{1}{n}
            \frac{n}{n+1}
            \biggl[
            -\biggl|y
            - \frac{1}{2}\biggr|^\frac{n+1}{n}
            + \biggl(\frac{1}{2}\biggr)^\frac{n+1}{n} 
            \biggr]
\end{equation}
Substituting the velocity profile into the flow rate constrain (Eq.\ \ref{Eq: Flow_rate_constraint}) gives the analytical relation between the pressure gradient for a Carreau fluid flowing between two plates in the limit of $\Cu\gg1$ 
\begin{equation} \label{Eq: dpdz_appendix}
	-\frac{d p}{d z}
	= 2^{n+1}
	\biggl[\frac{2n+1}{n}\biggr]^n
	\Cu^{n-1}.
\end{equation}
The scaled pressure gradient (Eq. \ref{Eq: P}) is then calculated by dividing Eq.\ (\ref{Eq: dpdz_appendix}) by the corresponding Newtonian pressure gradient that finally yields
\begin{equation} \label{Eq: dPdz_analyt_large}
	\mathcal{P}
	= \frac{-\dfrac{d p}{d z}}{\cancelto{12}{\biggl(-\dfrac{d p}{d z}\biggr)_{N}}}
	=  \frac{2^{n-1}}{3}
	\biggl[\frac{2n+1}{n}\biggr]^n
	\Cu^{n-1}
\end{equation}

\bibliography{Carreau_bibliography}

\end{document}